\documentclass[manuscript]{aastex6}
\usepackage{amsmath}
\usepackage{gensymb}
\usepackage{chngcntr}

\usepackage{graphicx}	
\usepackage{longtable}

\newcommand{\RNum}[1]{\uppercase\expandafter{\romannumeral #1\relax}}

\fullcollaborationName{The Friends of AASTeX Collaboration}

\begin{document}
\title{Spatial variation of the chemical properties of massive star-forming clumps}
\author{Mingyue Li\altaffilmark{1,4}, Jianjun Zhou\altaffilmark{2,3},
Jarken Esimbek\altaffilmark{2,3}, Donghui Quan\altaffilmark{2,5}, Yuxin He\altaffilmark{2,3}, Qiang Li\altaffilmark{2,4}, and Chunhua Zhu\altaffilmark{1}}
\altaffiltext{1}{School of Physical Science and Technology, Xinjiang University, Urumqi, 830046, P. R. China; limingyue@xao.ac.cn; chunhuazhu@sina.cn}
\altaffiltext{2}{Xinjiang Astronomical Observatory, Chinese Academy of Sciences, Urumqi 830011, P. R. China; zhoujj@xao.ac.cn}
\altaffiltext{3}{Key Laboratory of Radio Astronomy, Chinese Academy of Sciences, Urumqi 830011, P. R. China}
\altaffiltext{4}{University of the Chinese Academy of Sciences, Beijing 100080, P. R. China}
\altaffiltext{5}{Eastern Kentucky University, Richmond KY 40475, USA}

\begin{abstract}
  We selected 90 massive star forming clumps with strong N$_2$H$^+$(1-0), HCO$^+$(1-0), HCN(1-0) and HNC(1-0) emission from MALT90 survey. We obtained Herschel data for all 90 sources, and NVSS data for 51 of them. We convolved and regridded all images to the same resolution and pixel size, and derived the  temperature, H$_2$ column density, and molecules' abundances and abundance ratios of each pixel. Our analysis yields three main conclusions. Firstly, the abundances of N$_2$H$^+$, HCO$^+$, HCN and HNC increase when the column density decreases and the temperature increases, with spatial variations in their abundances dominated by changes in H$_2$ column density. Secondly, the abundance ratios between N$_2$H$^+$, HCO$^+$, HCN and HNC also display systemic variations as a function of column density due to the chemical properties of these molecules. Thirdly, the sources associated with the 20cm continuum emission can be classified into four types based on the behavior of the abundances of the four molecules considered here as a function of this emission. The variations of the first three types could also be attributed to the variation of H$_2$ column density.

\end{abstract}


\section{Introduction}
Many recent studies have focused on the chemistry of massive star forming regions at different evolutionary stages \citep{ger14,mie14,hoq13,sak08,sak10,sak12,san12,san13,vas11,vas12,zha16}. These studies did not always yield fully consistent results. \citet{ger14} and \citet {san12} suggested that N$_{2}$H$^{+}$/HCO$^{+}$ abundance ratio could serve as a chemical clock for massive star formation, whereas \citet{hoq13} found that the N$_{2}$H$^{+}$/HCO$^{+}$ abundance ratio shows no discernable trend from quiescent to protostellar, or to H\,{\sc{ii}}/PDR stage. The N$_{2}$H$^{+}$/HCO$^{+}$ abundance ratio of \citet{ger14} and \citet{san12} even displays different trend with the evolution.

Because these previous studies usually use the molecular data from single point observation, their results are probably affected by the distance. On the other hand, the chemical properties of star forming regions may be affected by the environment due to the fact that stars are formed in clusters. In two previous studies  \citep{han15,zha16}, we studied the global chemical properties of massive star forming regions at different evolution stages, and tried our best to remove the sources affected by the environment, and then obtained much improved results. However, it is still difficult to get an accurate chemical clock for massive star forming regions.

One possible reason may be that massive star forming regions are very complex, because stars are usually formed in clusters. Newly formed stars have effects on the physical and chemical conditions of the nearby molecular clouds. In fact, \citet{sch92} determined the HCN, HNC and DCN abundances and estimated the kinetic temperature at seventeen selected positions in the massive star forming region OMC-1. They found that the HCN/HNC abundance ratio is very high ($\sim$80) in the immediate vicinity of Orion-KL but declines rapidly in adjacent ridge positions down to values of $\sim$5. [H$^{13}$CN]/[HC$^{15}$N] ratio increases from 5-7 to roughly 15 close to Orion-IRc2. \citet{tfa06} obtained the radial profile of abundances for 13 molecules in small star forming regions for L1498 and L1517B. They found most abundance profiles can be described with simple step functions with a constant value in the outer layers and a central hole. However, such observational studies are still few to give a statistically reliable result.

As the chemistry of star-forming regions is very sensitive to prevailing physical conditions (temperature,
density, ionisation degree), understanding the chemical composition is of great importance to reveal the physics of the early stages of massive star formation. Spatial variations of chemistry reflect the response of the molecular emission to different physical and chemical environments characterizing the star forming clumps.

In this work, we select 90 massive star forming clumps with relatively strong molecular line emission (see Table \ref{tab1} and \ref{tab2}) from \citet{zha16} and study the spatial variations of the chemical properties of massive star forming regions. We propose that good chemical clock can not be achieved from either the data of single point observation or the data averaged from the whole source. We describe our archive data in Section \ref{sec2}, data reduction and results in Section \ref{sec3}. We discuss the spatial variations of the abundances and abundance ratios with temperature and column density, the effect of 20\,cm continuum emission on the abundances and abundance ratios in section \ref{sec4}. We summarize our findings in section \ref{sec5}. The captions of 12 figures online are listed in Appendix.

 \begin{table*}
\tiny
 \centering
  \begin{minipage}{155mm}
   \caption{Our sample includes 90 massive star forming clumps selected from \citet{zha16}. Here are 51 sources which have more than 10 pixels overlapped with the 20\,cm continuum emission (see details in Section 4.5), they are numbered from 1 to 51. For the column ``type", First, Second, Third and Fourth designate the different correlation between the 20cm continuum emission and abundances of N$_2$H$^+$, HCO$^+$, HCN, and HNC (see details in Section 4.5). For the column ``Group", I and II designate the sources overlapped with the bubble and without the bubble, respectively. Their evolution stages were determined by \citet{zha16}.}\label{tab1}
    \begin{tabular}{ccccc}
        \hline
        \multicolumn{1}{c}{No.} & \multicolumn{1}{c}{Name} & \multicolumn{1}{c}{Type}  & \multicolumn{1}{c}{Group} & \multicolumn{1}{c}{Evolution stage} \\
  \hline
  1	&   G000.006+00.156  & Third    &  II  &   C   \\
  2	&   G000.410-00.504  & Fourth  &  II  &   C   \\
  3	&   G000.633+00.601  & Fourth &   C   &  I  \\
  4    &   G005.637+00.238  & Fourth   &  II  &   C \\
  5    &   G005.831-00.512  & Second  &  II     &   C \\
  6    &   G006.119-00.636   & Fourth    &  II &   C   \\
  7    &   G006.216-00.609  & Fourth    &  II   &   B  \\
  8    &   G006.796-00.256   & First   &  I  &   C  \\
  9   &   G008.350-00.317   & Fourth   &  II  &   C \\
  10   &   G009.620+00.195   & First  &  II &   C   \\
  11   &   G010.624-00.383    & Third  &  I   &   C  \\
  12   &   G011.112-00.399    & First    &  I  &   C  \\
  13   &   G012.200-00.033   & Fourth    &  I  &   B  \\
  14   &   G012.497-00.222  & Second   &  I   &   B   \\
  15   &   G012.774+00.337   & Fourth   &  II  &   C  \\
  16   &   G013.209-00.141    & Fourth  &  II   &   C  \\
        \hline
     \end{tabular}
     \medskip
   \end{minipage}
\end{table*}

\setcounter{table}{0}
\begin{table*}
\scriptsize
 \centering
  \begin{minipage}{155mm}
   \caption{continued.}
    \begin{tabular}{ccccc}
        \hline
        \multicolumn{1}{c}{No.} & \multicolumn{1}{c}{Name} & \multicolumn{1}{c}{Type}  & \multicolumn{1}{c}{Group} & \multicolumn{1}{c}{Evolution stage} \\
         \hline
  17   &   G014.226-00.511    & Fourth  &  I  &   C \\
  18   &   G014.777-00.486   & Second  &  I   &   B  \\
  19   &   G347.967-00.434 &   First   &  II   &  C \\
  20   &   G348.228+00.413 & Second   &  I   & C \\
  21   &   G349.137+00.024    & First    &  II &   C \\
  22   &   G350.111+00.092  & First   &  II   &   C   \\
  23   &   G350.412-00.062   & Third  &  II  &   A   \\
  24   &   G350.506+00.958  & Third  &  II   &   C  \\
  25   &   G350.522-00.349  & Fourth   &  II  &   C  \\
  26   &   G350.687-00.491   & Second   &  I  &   C \\
  27   &   G350.710+01.027   & Second   &  I  &   C \\
  28   &   G350.763+00.793  & Fourth  &  II  &   C  \\
  29   &   G351.040-00.336  & Third  &  II   &   C   \\
  30   &   G352.072+00.679   & Fourth  &  I   & B   \\
  31   &   G352.233-00.162  & Fourth  &  II  &   C  \\
  32   &   G352.315-00.443  & Third   &  I  &   C   \\
  33   &   G352.492+00.796 & Second   &  II  &  C  \\
  34   &   G352.684-00.120 & Second   &  I   &  C  \\
  35   &   G352.857-00.203 & Third   &  I   &  C    \\
  36   &   G352.972+00.925  & Third   &  II   &  C \\
  37   &   G353.010+00.983 & Fourth   &  I   &  C  \\
  38   &   G353.115+00.366 & First    &  II    &  C \\

      \hline
     \end{tabular}
     \medskip
   \end{minipage}
\end{table*}

\setcounter{table}{0}
\begin{table*}
\scriptsize
 \centering
  \begin{minipage}{155mm}
   \caption{continued.}
    \begin{tabular}{ccccc}
        \hline
        \multicolumn{1}{c}{No.} & \multicolumn{1}{c}{Name} & \multicolumn{1}{c}{Type}  & \multicolumn{1}{c}{Group}  & \multicolumn{1}{c}{Evolution stage} \\
         \hline
   39   &   G353.147+00.851 & Second   &  I   &  C    \\
  40   &   G353.198+00.927   & First &  II  &  C  \\
  41  &   G353.271+00.641   & Third &  II &  C   \\
  42   &   G353.462+00.563  & Fourth   &  I  &  B    \\
  43   &   G353.975+00.256  & Fourth  &  I  &  B    \\
  44   &   G355.265-00.269  & Third    &  I  &  B  \\
  45   &   G355.740+00.655  & Second  &  I   &  B  \\
  46   &   G355.829-00.501 & Fourth   &  I   &  C \\
  47   &   G356.482+00.190 & Second  &  I   &  B   \\
  48   &   G356.517+00.664 &  Fourth   &  I  &  B  \\
  49   &   G358.388-00.484  & Fourth &  I  &  C   \\
  50   &   G359.716-00.375  & Fourth   &  II  &  C  \\
  51   &   G359.911-00.305 & Second   &  I  &  B   \\
      \hline
     \end{tabular}
     \medskip
   \end{minipage}
\end{table*}

\begin{table*}
\tiny
 \centering
  \begin{minipage}{155mm}
   \caption{Our sample includes 90 massive star forming clumps selected from \citet{zha16}. Here are the remaining 39 sources, they are numbered from 52 to 90. For the column ``Group", I and II designate the sources overlapped with the bubble and without the bubble, respectively. Their evolution stages were determined by \citet{zha16}. }\label{tab2}
    \begin{tabular}{cccc}
        \hline
        \multicolumn{1}{c}{No.} & \multicolumn{1}{c}{Name} & \multicolumn{1}{c}{Evolution stage}  & \multicolumn{1}{c}{Group} \\
  \hline
  52    &   G005.893-00.320 &   C   &  II \\
  53   &   G010.724-00.332 &   C   &  II \\
  54   &   G011.942-00.156 &   C   &  II   \\
  55   &   G012.418+00.506 &   C   &  II  \\
  56   &   G013.657-00.599 &   C   &  I  \\
    57   &   G340.054-00.244 &   C   &  I   \\
  58   &   G340.104-00.313 &   C   &  II   \\
  59   &   G340.229-00.144 &   A   &  I     \\
  60   &   G340.311-00.436 &   B   &  I     \\
  61   &   G340.632-00.648 &   B   &  II   \\
  62   &   G340.785-00.097 &   B   &  II  \\
  63   &   G341.034-00.114 &   A   &  I   \\
  64   &   G341.127-00.350 &   C   &  II \\
  65   &   G342.484+00.183 &   C   &  I    \\
  66   &   G342.706+00.125 &   C   &  II  \\
  67   &   G342.822+00.382 &   B   &  I   \\
  68   &   G343.127-00.063 &   B   &  II   \\
  69   &   G343.520-00.519 &   B   &  II   \\
  70   &   G343.738-00.112 &   B   &  I  \\
        \hline
     \end{tabular}
     \medskip
   \end{minipage}
\end{table*}

\setcounter{table}{1}
\begin{table*}
\scriptsize
 \centering
  \begin{minipage}{155mm}
   \caption{continued.}
    \begin{tabular}{cccc}
        \hline
        \multicolumn{1}{c}{No.} & \multicolumn{1}{c}{Name} & \multicolumn{1}{c}{Evolution stage}  & \multicolumn{1}{c}{Group}  \\
         \hline
  71   &   G343.780-00.235 &   B   &  I     \\
  72   &   G344.915-00.229 &   B   &  I    \\
  73   &   G345.261-00.418 &   C   &  I    \\
  74   &   G346.078-00.056 &   C   &  II   \\
  75   &   G346.307+00.114 &   C   &  II   \\
  76   &   G346.369-00.648 &   B   &  I  \\
  77   &   G350.271-00.500 &   A   &  I   \\
     78   &   G352.060+00.603 &   A   &  I \\
  79   &   G352.142-01.016 &   B   &  I   \\
  80   &   G353.577+00.661 &  B   &  I \\
  81   &   G354.628-00.610 &  A   &  I   \\
  82   &   G354.945-00.539 &  C   &  I   \\
  83   &   G355.182-00.419 &  B   &  I   \\
  84   &   G355.412+00.103 &  B   &  I   \\
  85   &   G356.008-00.424 &  B   &  I  \\
  86   &   G356.008-00.758 &  C   &  II   \\
  87   &   G356.255-00.056 &  C   &  I   \\
  88   &   G356.344-00.068 &  B   &  I    \\
  89   &   G357.554-00.550 &  C   &  I   \\
  90   &   G358.980+00.083 &  B   &  I   \\
        \hline
     \end{tabular}
     \medskip
   \end{minipage}
\end{table*}

\section{Archive data}\label{sec2}

\subsection{The ATLASGAL survey}
The ATLASGAL Survey was the first systematic survey of the inner Galactic plane in the submm band traces the thermal emission from dense clumps at 870 $\mu$m and is complete to all massive clumps above 1000M$_{\odot}$ to the far side of the inner Galaxy ($\sim$20 kpc). The survey was carried out with the Large
\emph{APEX} Bolometer Camera \citep{sch09}, which is an array of 295 bolometers observing at 870 $\mu$m (345 GHz). At this wavelength, the \emph{APEX} Telescope has a full width at half maximum (FWHM) beam size of 19.2 arcsec. The survey region covered the Galactic longitude region of $|\ell|<60^{\circ}$
and 280$^{\circ}$ $<$ $\ell$ $<$ 300$^{\circ}$, and Galactic latitude $|$b$|$ $<$ 1.5$^{\circ}$ and $-$2$^{\circ}$ $<$ $\ell$ $<$ 1$^{\circ}$, respectively.
\citet{urq14} presented a compact source catalogue of this survey, which consists of $\sim$10163 sources and is 99 per cent complete at a $\sim$6$\sigma$ flux level, which corresponds to a flux sensitivity of $\sim$0.3-0.4 Jy beam$^{-1}$.

\subsection{The MALT90 survey}
The MALT90 survey is a large international project that exploited the fast-mapping capability of the ATNF Mopra 22m telescope and obtained 16 molecular line maps near 90 GHz in order to characterize the physical and chemical conditions of high-mass
star formation regions over a wide range of evolutionary states (from Pre-stellar cores, to Proto-stellar cores, and to H II regions). The sample of this survey is a sub-sample of the ATLASGAL catalog. The angular and spectral resolution of this survey are about 36$^{\prime\prime}$ and 0.11 km s$^{-1}$ \citep{jac13}. The MALT90 data has been obtained from the
online archive\footnote{http://atoa.atnf.csiro.au/MALT90/}.

 \subsection{the NVSS survey}
 The NRAO VLA Sky Survey (NVSS) is the  continuum survey, which covers the entire sky north of 40$^{\circ}$ declination at 1.4 GHz. The angular resolution is about 45$^{\prime\prime}$, and the noise level in the images is about 0.5 mJy/beam. A detailed description of the survey can be found in \citet{con98}. In this paper, we use images provided by the home page of NVSS\footnote{http://www.cv.nrao.edu/nvss/}.

\subsection{The GLIMPSE and MIPS survey}
The Galactic Legacy Infrared Mid-Plane Survey Extraordinaire (GLIMPSE) survey is a mid-infrared survey (3.6, 4.5, 5.8, and 8.0 $\mu$m) of the Inner Galaxy performed with the \emph{Spitzer} Space Telescope \citep{ben03}. The angular resolution is better than 2$^{\prime\prime}$ at all wavelengths. GLIMPSE covers $5^{\circ}\leq|\ell|\leq65^{\circ}$ with $|b|\leq1^{\circ}$, $2\leq|\ell|<5^{\circ}$ with $|b|\leq1.5^{\circ}$, and $|\ell|<2^{\circ}$ with $|b|\leq2^{\circ}$. The MIPS/\emph{Spitzer} Survey of the Galactic Plane (MIPSGAL) is a survey of the same region as GLIMPSE at 24 and 70 $\mu$m, using the Multiband Imaging Photometer (MIPS) aboard the \emph{Spitzer} Space Telescope \citep{rie04}. The angular resolution at 24 and 70 $\mu$m is 6$^{\prime\prime}$ and 18$^{\prime\prime}$, respectively.

\subsection{The Hi-GAL survey}
The {\it{Herschel}} Infrared Galactic (Hi-GAL) Plane Survey is an Open Time Key Project on-board the \textit{Herschel Space Observatory} \citep{pil10}, which mapped the Inner Galactic at 70 and 160 $\mu$m with Photoconductor Array Camera and Spectrometer (PACS, \citet{pog10}) and 250, 350, and 500 $\mu$m with the Spectral and Photometric Imaging Receiver (SPIRE, \citet{gri10}). The spatial resolution of the images are 6$\arcsec$, 12$\arcsec$, 18$\arcsec$, 24$\arcsec$, and 35$\arcsec$ for the five wavelength bands \citep{mol10}, respectively.

Finally, we summary the character properties of above 5 surveys and what purposes we use these data in this work in Table \ref{tab3}.

\begin{table*}
\scriptsize
 \centering
  \begin{minipage}{155mm}
   \caption{Here we present a summary of five surveys introduced in Section 2 and our purposes used them in this work. We list the project name in column 1, frequency or wavelength in column 2, facility in column 3, purpose using the corresponding data in column 4 and citation in column 5. }\label{tab3}
    \begin{tabular}{lllll}
        \hline
        \multicolumn{1}{c}{Survey} & \multicolumn{1}{c}{Frequency (Wavelength)} & Facility  & \multicolumn{1}{c}{Purpose here} & \multicolumn{1}{c}{Citation}  \\
         \hline
  ATLASGAL   &   345\,GHz &  APEX & H$_2$ column density   &  \citet{sch09}  \\
  MALT90	   &   90\,GHz &  Mopra   & Molecules' abundances   &  \citet{jac13}  \\
  NVSS          &   1.4\,GHz &  VLA    & Tracer of UV radiation   & \citet{con98}\\
  GLIMPSE   &   3.6, 4.5, 5.8 and 8.0\,$\mu$m & Spitzer  &  Tracer of star formation  & \citet{ben03} \\
  MIPS    &   24 and 70\,$\mu$m &  Spitzer & Tracer of star formation   &  \citet{rie04}  \\
  Hi-GAL    &   70, 160, 250, 350, 500\,$\mu$m & Herschel  &  Dust temperature &\citet{mol10}  \\
      \hline
     \end{tabular}
     \medskip
   \end{minipage}
\end{table*}

\begin{figure}
\figurenum{1}
  \begin{center}
  \includegraphics[width=0.5\textwidth]{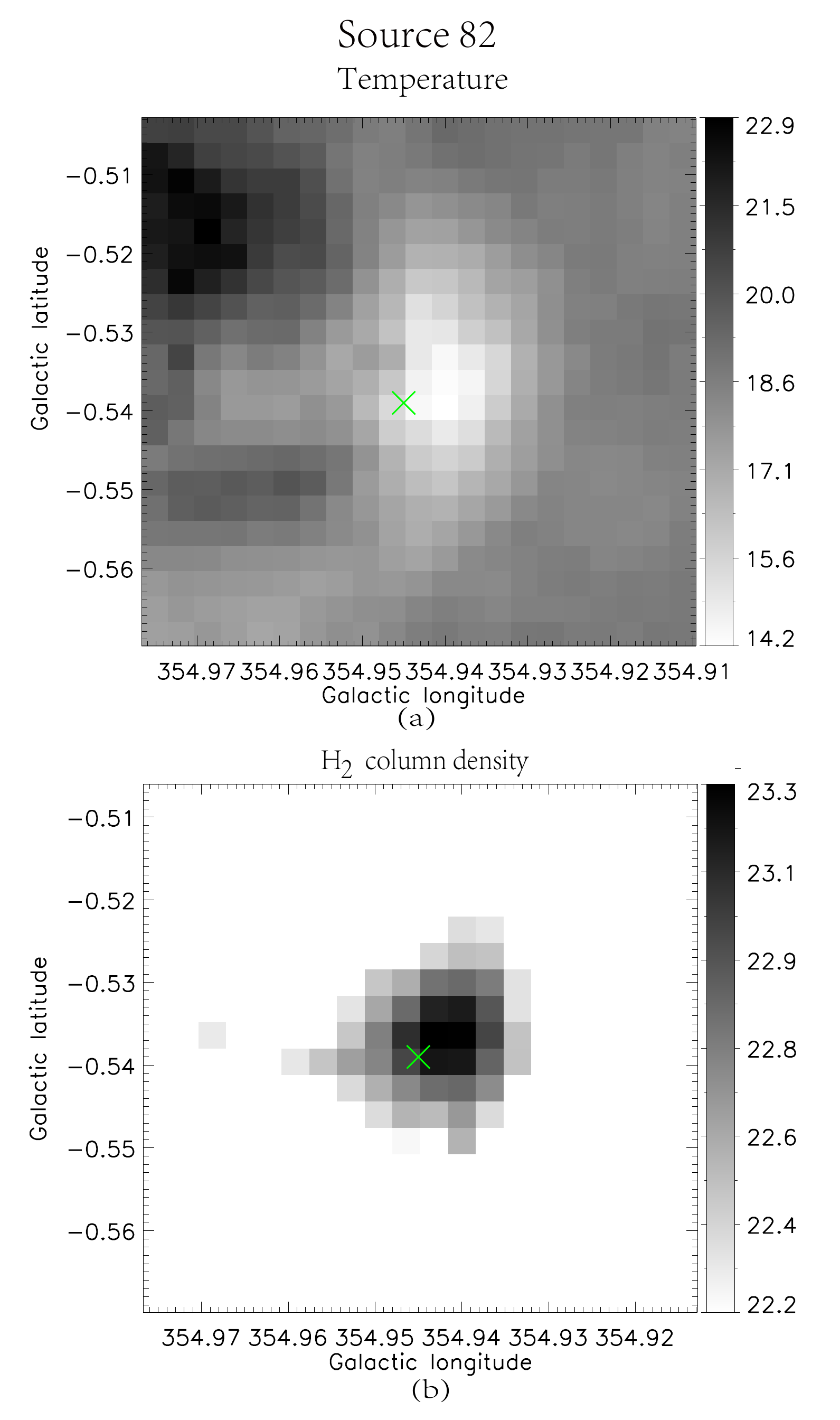}
  \end{center}
  \caption{The distributions of temperature (panel a) and H$_2$ column density (panel b) of the source 82 (G354.945-00.539) in gray scale. The black bars show temperature in K and the logarithm of column density in cm$^{-2}$.} \label{fig1}
\end{figure}

\begin{figure}
\figurenum{2}
  \begin{center}
  \includegraphics[width=0.5\textwidth]{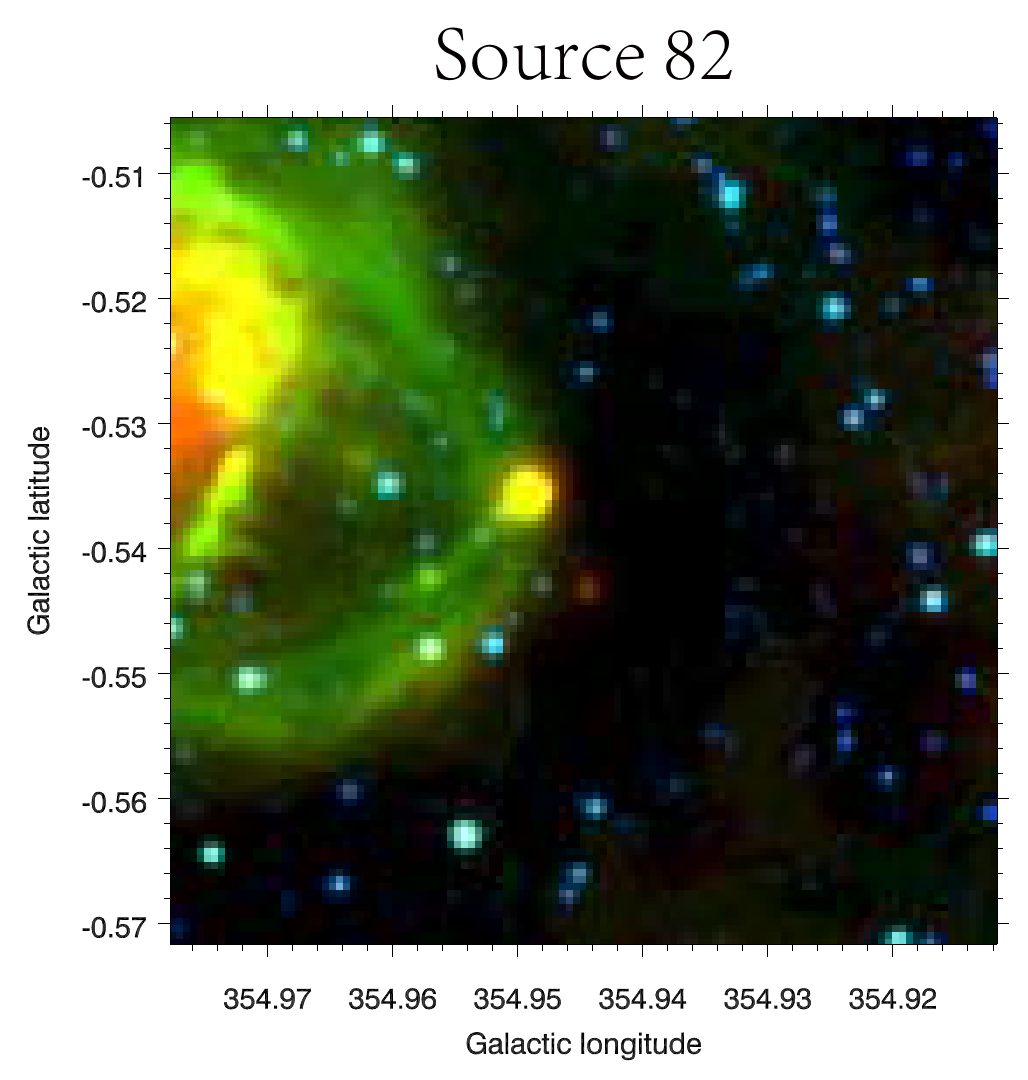}
  \end{center}
  \caption{Three-color images of the source 82. We show 3.6$\mu$m in blue, 8$\mu$m in green and 24$\mu$m in red.} \label{fig2}
\end{figure}

\begin{figure}
\figurenum{3}
  \begin{center}
  \includegraphics[width=0.95\textwidth]{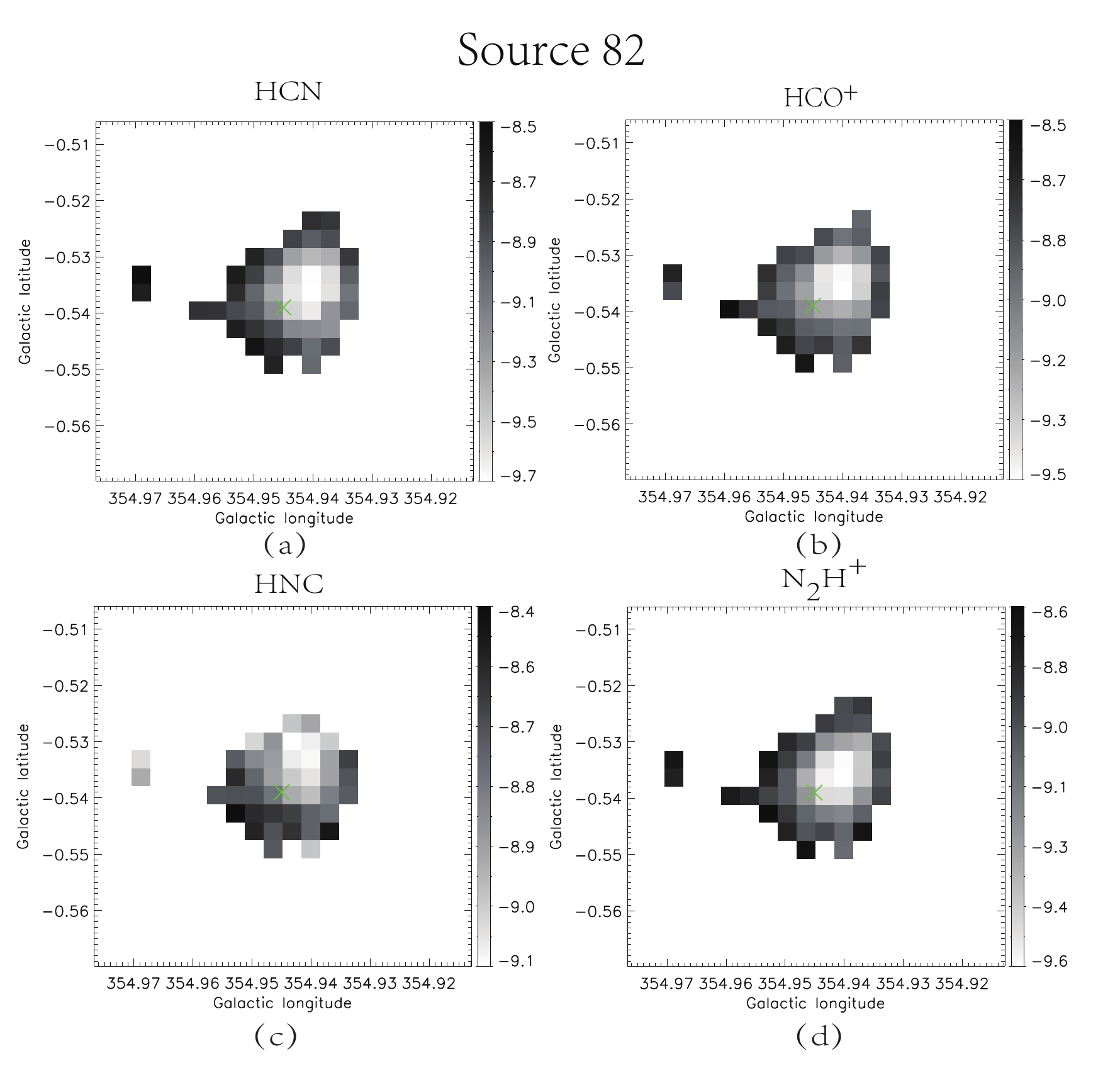}
  \end{center}
  \caption{The abundance distributions of N$_{2}$H$^{+}$, HCO$^{+}$, HCN and HNC of the source 82 (G354,945-00.539) in gray scale. The black bars show the abundance values in log.}\label{fig3}
\end{figure}

\section{Results}\label{sec3}
\subsection{The distribution of dust temperature and column density}\label{sec3.1}
 Assuming the dust emission is optically thin for the Herschel data at 160, 250, 350, and 500 $\mu$m, we use these data to derive the distributions of temperature and column density of 90 massive star forming clumps. To obtain the distribution of the dust temperature, we convolved and regridded the 160, 250, and 350 $\mu$m images to the lowest resolution 35$\arcsec$ and
largest pixel size 11$\arcsec$.5 of the 500 $\mu$m image, and then performed SED fitting pixel-by-pixel. We used a grey body function for a single temperature as following \citep{War90} to estimate the temperature of each pixel.
 \begin{equation}\label{eq-1}
          F_{\nu} = \Omega B_{\nu}(T_{\mathrm{d}})(1-e^{-\tau}),
          \end{equation}
where $\Omega$ is the effective solid angle corresponding to each pixel, $B_{\nu}(T_{\mathrm{d}})$ is the
Plank function at the dust temperature $T_{\mathrm{d}}$, which is defined as $B_{\nu}(T_{\mathrm{d}}) =
(2h\nu^{3}/ c^{2})[\mathrm{exp}(h\nu/kT)-1]^{-1}$, where $\textit{h}$ is the Planck constant, $\nu$ is the
frequency, $\textit{c}$ is the speed of light, and $\textit{k}$ is the Boltzmann constant.
The optical depth $\tau$ is given by the relation $\tau = (\nu/ \nu_{c})^{\beta}$, where $\beta = 2$  is the
dust emissivity index, and $\nu_{c}$ is the critical frequency at which $\tau = 1$. We assigned an uncertainty of 20$\%$ to each \textit{Herschel} flux and used it as a calibration error \citep{fai12}.

 We display the distributions of temperature for 90 sources in Figure 1A (see an electronic version online).

Subsequently the ATLASGAL 870 $\mu$m images were also convolved and regridded to the resolution and
pixel size of the 500 $\mu$m image.
The hydrogen molecular column density distribution of each source was then calculated pixel by pixel:
              \begin{equation}\label{eq-2}
               N_{\mathrm{H}_{2}}={{S_{\nu}}R\over{B_\nu}(T_{\mathrm{d}}){\Omega}{\kappa_{\nu}}{\mu}{m_{\mathrm{H}}}},
              \end{equation}
where S$_{\nu}$ is the total 870 micron flux density over the line-emitting area; R is the gas-to-dust mass ratio, which is assumed to be 100; $\Omega$ is the solid angle; $\mu$ is the mean molecular weight of the interstellar medium, which is assumed to be 2.8; m$_{H}$ is the mass of a hydrogen atom; B$_{\nu}$ is the Planck function for dust temperature T$_{D}$; and $\kappa$$_{\nu}$ is the dust-absorption coefficient, taken as 1.85 cm$^{2}$ g$^{-1}$ (interpolated to 870 $\mu$m from column 5 of
Table 1 of \citet{oss94}).

We thus obtained the distributions of H$_2$ column density for 90 sources, and shown them in Figure 2A (see an electronic version online).

 An examination of temperature and H$_2$ column density distributions (Figure \ref{fig1}) and the corresponding three-color images (Figure \ref{fig2}, in Figure 3A online we show the three-color images from Spitzer data at 3.6, 8 and 24$\mu$m for all 90 sources), we found that temperature is usually higher at the center of the clump where H$_2$ column density is high for the sources in which there is star formation. Conversely, the temperature is lower at the center of the clump where no stars are being formed. Figure \ref{fig1} shows the temperature and column density distributions of the source 82 as an example. It is clear that the temperature is lower when the column density is higher. This indicates that the more diffuse outer regions of the clump may be heated by interstellar UV radiation and cosmic rays. It is hot where the extinction is small. The very high temperature regions in the upper-left corner of Figure  \ref{fig1}(a) suggests the presence of an infrared bubble.

\subsection{The distribution of abundances and abundance ratios}\label{sec3.2}
 Following \citet{san12}, we derived the optical depth of each pixel using the formula
 \begin{equation}
 T_{\rm mb} = f[J(T_{\rm ex}) - J(T_{\rm bg})](1 - e^{-\tau_{\nu}})~,
 \label{Tmb-eq}
 \end{equation}
 where $T_{\rm mb}$ is the main beam brightness temperature, $f$ is the filling factor, $\tau_{\nu}$ is the optical depth of the line, $T_{\rm bg}$ is the background temperature, and $J(T) = \frac{h\nu}{k}\frac{1}{e^{h\nu/kT} - 1}~$, $h$ is the Planck constant, $k$ is the Boltzmann constant. To determine the optical depth of the line, the excitation temperature($T_{\rm ex}$), of the line was assumed to be equal to the dust temperature ($T_{\rm D}$).

 Then, the column density of each pixel was calculated based on the averaged molecular spectra of N$_{2}$H$^{+}$, HCO$^{+}$, HCN, and HNC of each pixel by assuming local thermodynamic equilibrium (LTE), using the following formula from \citet{gar91}
 \begin{eqnarray}
 N=\frac{3k}{8\pi^3B\mu^2R}\frac{(T_{\rm ex}+hB/3k)}{(J+1)}\frac{\exp(E_{_{J}}/kT_{\rm ex})}{[1-\exp(-h\nu/kT_{\rm ex})]} \nonumber
 \int\tau_{\nu}\,dv~,
 \label{eqn-den-colum}
 \end{eqnarray}

 where $\nu$ is the transition frequency and is assumed as 1, $\tau_{\nu}$ is the optical depth of the line, $\mu$ is the permanent dipole moment of the molecule, and $R$ is the relative intensity of the brightest hyperfine transition with respect to the others. $R$ is only relevant for hyperfine transitions because it considers the satellite lines corrected by their relative opacities. $R = 5/9$ for N$_2$H$^+$, 1 for HCO$^{+}$ and HNC, and 3/5 for HCN. $J$ is the rotational quantum number of the lower state, $E_J=hBJ(J+1)$ is the energy in level $J$, and $B$ is the rotational constant of the linear molecule in question. The dipole moment and rotational constant of these four molecules are shown in Table 3 of \citet{zha16}.

 We thus derived the column densities of N$_{2}$H$^{+}$, HCO$^{+}$, HCN and HNC, and their abundance distributions for all 90 sources. We display the abundance distributions for all 90 sources in Figure 4A (see an electronic version online). We also calculated the abundance ratios between these four molecules for each source, and display their distributions in Figure 5A (see an electronic version online).

\begin{figure}
\figurenum{4}
  \begin{center}
  \includegraphics[width=0.95\textwidth]{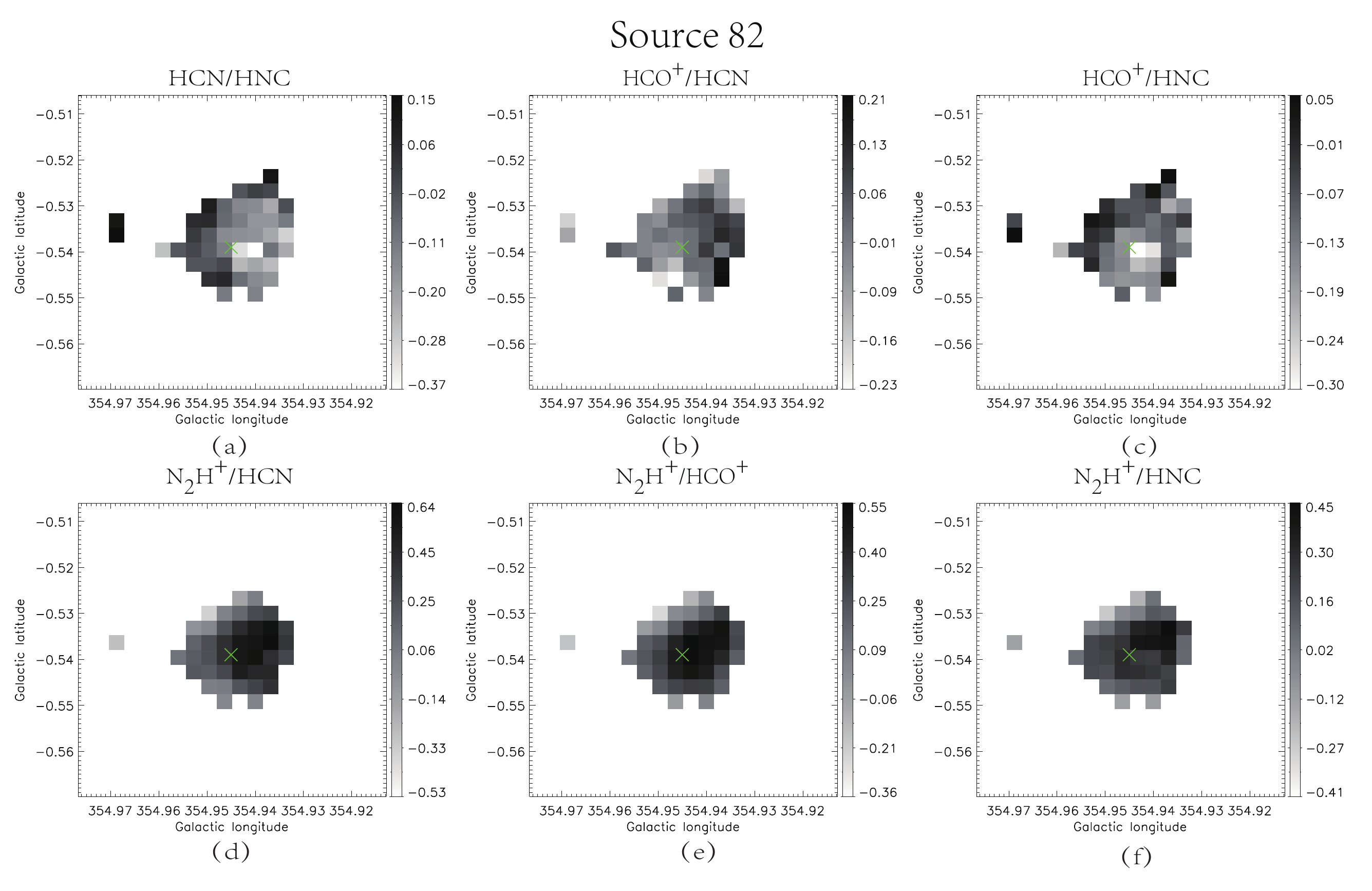}
  \end{center}
  \caption{The distributions of abundance ratio between N$_{2}$H$^{+}$, HCO$^{+}$, HCN and HNC for the source 82 (G354.945-00.539) in gray scale. The black bars show the units of in log.}\label{fig4}
\end{figure}

 The results in Figure 4A and 5A indicate that spatial variations of abundance and abundance ratio in massive star forming clumps are complex, making it difficult to describe them clearly in a unified way. As an example, Figure \ref{fig3} shows the abundance distributions of N$_{2}$H$^{+}$, HCO$^{+}$, HCN and HNC for the source 82. By comparing Figure \ref{fig3} with Figure \ref{fig1}, we see clearly the abundances of these four molecules are usually higher at edges of the clump where the density is lower and temperature is higher, and are usually lower at the center of the clump where the conditions are opposite. However, the spatial variations of the abundance ratios of these molecules seem to be complex (see Figure \ref{fig4}). The abundance ratio of HCN/HNC is large in eastern edge of the clump,  and it seems decreasing from the eastern edge to the western edge of the clump. By contrast, the abundance ratio of HCO$^{+}$/HCN is small in the eastern edge of the clump, and increases from the eastern edge to the western edge of the clump. The abundance ratio of HCO$^{+}$/HNC is large in the eastern, northern and western edges of the clump, and is small in the center and southern part of the clump. These results suggest both HCN and HCO$^+$ abundances increase to the eastern edge of the clump, with HCN increasing more rapidly than HCO$^+$. Conversely, HNC abundance may decrease in the eastern edge. The abundance ratios of N$_{2}$H$^{+}$/HCN, N$_{2}$H$^{+}$/ HCO$^{+}$, N$_{2}$H$^{+}$/HNC display similar spatial variations, with ratios being large in the center and western part of the clump. The results support the idea that  HCN and HCO$^+$ abundances increase rapidly in the eastern edge, and may also indicate that N$_{2}$H$^{+}$ abundance decreases in the eastern edge, more rapidly than HNC abundance does.

\section{Discussion}\label{sec4}
\subsection{The chemical properties of N$_2$H$^+$, HCO$^+$, HCN and HNC}\label{sec4.1}
\subsubsection{HCN and HNC}\label{sec4.1.1}
Past observations show that HCN is a widely used dense gas tracer, and HNC is a reliable tracer of cold gas \citep{sch92}. The abundances of CN, HCN, and HNC are sensitive to temperature, optical depth, and UV radiation \citep{sch92,fue93,bog05}. HNC transfers to HCN at high temperature. Therefore the abundance ratio of HCN/HNC is strongly dependent on the temperature \citep{vas11,ung97}. The abundance ratio HCN/HNC is $\sim$1 for infrared dark clouds (IRDCs; Vasyunina et al. 2011), and is higher and reaches up to $\sim$13 for high-mass protostars (HMPOs; \citet{hel97}), and is $\sim$80 for very warm regions like Orion \citep{sch92,gol86}.  Gas-phase chemical models suggest that HCN and HNC are primarily produced via the dissociative recombination reaction \citep{her78}
 \\HCNH$^+$ +e$^-$ $\rightarrow$ HCN + H or HNC + H.
  \\The resulting HNC/HCN abundance ratio is predicted to be 0.9 in this case. HNC has a unique formation via the reactions \citep{pea74,all80}
  \\H$_2$CN$^+$/H$_2$NC$^+$+ e$^-$$\rightarrow$ HNC + H.
  \\As a result of this additional HNC production channel, the HNC/HCN ratio can rise above unity \citep{mie14}.

\subsubsection{N$_2$H$^+$}\label{sec4.1.2}
Owing to its resistance to depletion at low temperatures and high densities,  N$_2$H$^+$ is an excellent tracer of cold and dense molecular clouds. It is primarily formed through the gas-phase reaction
\\ H$_3$$^+$ +N$_2$ $\rightarrow$ N$_2$H$^+$ +H$_2$. \\
When there are CO molecules in the gas phase, they can destroy N$_2$H$^+$, and produce HCO$^+$ through reaction \citep{mie14}
\\ N$_2$H$^+$ + CO $\rightarrow$ HCO$^+$ + N$_2$.

\subsubsection{HCO$^+$}\label{sec4.1.3}
 HCO$^+$  is a highly abundant molecule, with its abundance especially enhanced around regions of higher fractional ionization. It can also be enhanced by the presence of outflows where shock-generated radiation fields are present \citep{vas11}. In dense molecular clouds, HCO$^+$  is mainly formed through the gas-phase ion-neutral reaction \citep{her73}
 \\H$_3$$^+$ + CO $\rightarrow$ HCO$^+$ + H$_2$.
  \\When the shock heats the gas and produces UV radiation through Ly-$\alpha$ emission ($\lambda$ = 121.6 nm), the icy grain mantles evaporate and the HCO$^+$ abundance is enhanced because of evaporated CO and H$_2$O, which can form HCO$^+$ in the reaction with photoionised carbon \citep{raw00,raw04}
 \\C$^+$ + H$_2$O $\rightarrow$ HCO$^+$ + H.

\begin{figure}
\figurenum{5}
  \begin{center}
  \includegraphics[width=0.95\textwidth]{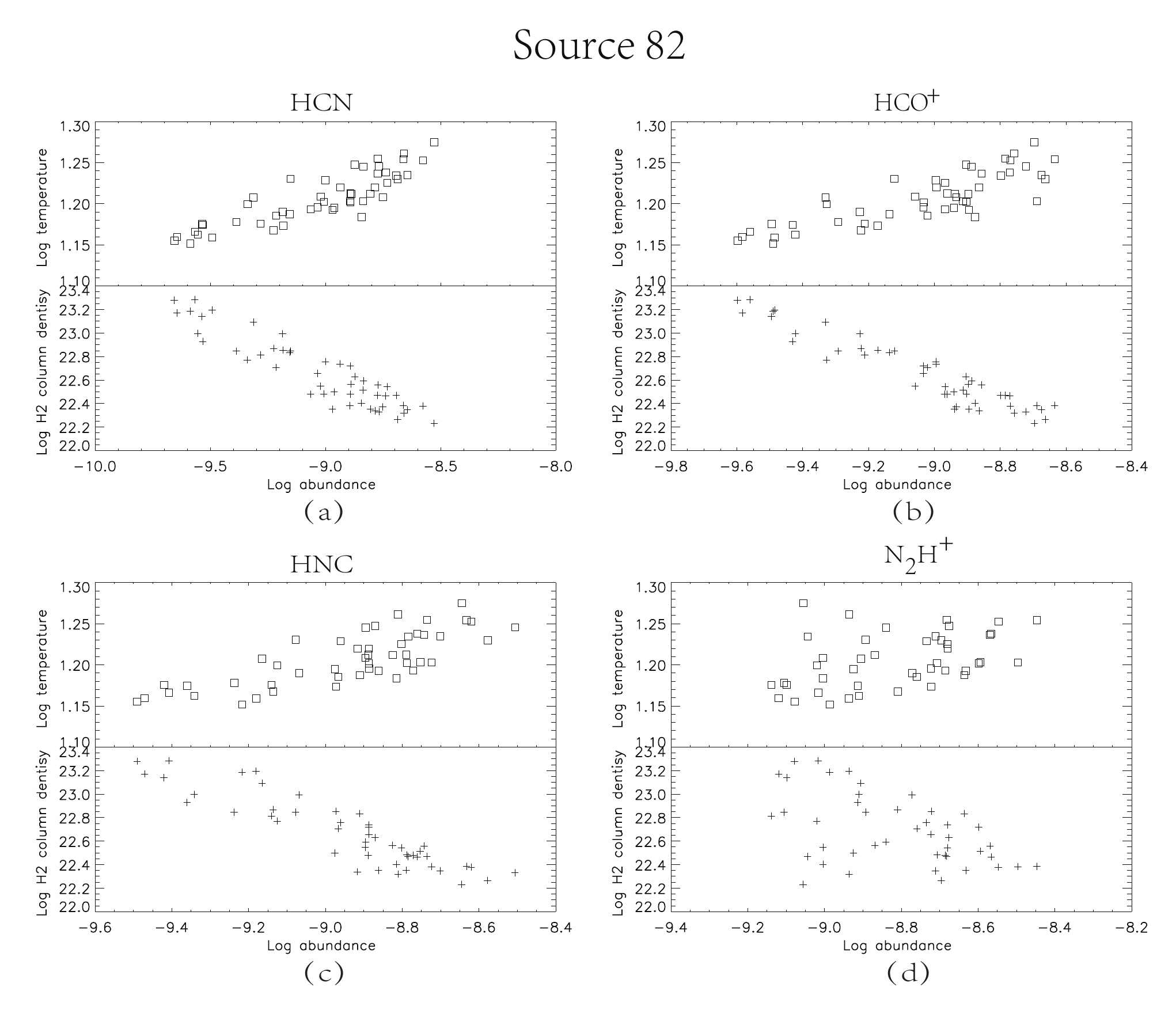}
  \end{center}
  \caption{The abundances of N$_2$H$^+$, HCO$^+$, HCN, HNC as a function of H$_2$ column density and temperature for the source 82 (G354.945-0.539), respectively. This source is an example of Group I.}\label{fig5}
\end{figure}

\begin{table*}
\scriptsize
 \centering
  \begin{minipage}{155mm}
   \caption{We classified 90 sources into Group I and II (see details in Section \ref{sec4.2}). Here we display the variation trend of the molecules' abundances with H$_2$ column density and temperature for Group I and II sources respectively. }\label{tab4}
    \begin{tabular}{lllll}
        \hline
        \multicolumn{1}{c}{Type} & \multicolumn{1}{c}{Environment} & \multicolumn{1}{c}{Abundance} & Abundance  \\
         \multicolumn{1}{c}{} & \multicolumn{1}{c}{}& \multicolumn{1}{c}{with N(H$_2$)} & with T  \\
         \hline
  Group I     & isolated &  decreasing (obvious) & increasing (weak)   \\
  Group II   &  near bubbles & decreasing (obvious)  & increasing (weak)   \\
      \hline
     \end{tabular}
     \medskip
   \end{minipage}
\end{table*}

\subsection{The spatial variations of the abundances}\label{sec4.2}
We studied the abundance variations of  N$_2$H$^+$, HCO$^+$, HCN and HNC as a function of column density and temperature pixel by pixel. Figure 7A online shows the results of all 90 sources. In general, the temperature ranges from 16 to 26 K, the column density ranges from 10$^{22}$ to 10$^{24}$ cm$^{-2}$. The abundances of N$_2$H$^+$, HCO$^+$, HCN and HNC range from 10$^{-10}$ to 10$^{-8}$, depending on the source. Abundances of N$_2$H$^+$, HCO$^+$, HCN and HNC often show an obvious increasing trend with the decreasing column density,  and a relatively weak but also obvious increasing trend with the increasing temperature.

 By checking the three color images of 90 sources (Figure 3A), we found out that some sources overlap with the bubbles along the line of sight. These bubbles are usually large with size similar to that of clumps, and show strong 8 and 24$\mu$m emission.The abundances of  N$_2$H$^+$, HCO$^+$, HCN and HNC of these sources usually display an obvious increasing trend as a function of decreasing H$_2$ column density, but with relatively larger dispersion. We classified these sources into Group II (see Table \ref{tab1} and \ref{tab2}). Other sources, however, are relatively isolated star forming regions. Some contain small H\,{\sc{ii}} or infrared bubble excited by their own star formation. The abundances of  N$_2$H$^+$, HCO$^+$, HCN and HNC clearly rise when H$_2$ column density decreases, and with smaller dispersion. We classified these sources into Group I (see Table \ref{tab1} and \ref{tab2}). We summary the abundance variation of Group I and II sources with H$_2$ column density and temperature in Table \ref{tab4}.

We display the source 82 as an example of Group I in Figure \ref{fig5}. It can be clearly seen the abundances of N$_2$H$^+$, HCO$^+$, HCN and HNC display an obvious increasing trend with the decreasing column density and increasing temperature.  We find that (1) H$_2$ column density is inversely proportional to the temperature, the pixels with higher column density tend to have lower temperature. (2) Abundances of these four molecules reach the minimum in the pixels with lowest temperature and largest column density.  (3) The abundance of N$_2$H$^+$ display a weak increasing trend with the decreasing column density and increasing temperature.

It should be noted that the column densities of N$_2$H$^+$, HCO$^+$, HCN, HNC measured at pixels with different column densities do not vary much when the H$_2$ column density vary about one order of magnitude (see Figure 6A online). This could be attributed to the relative large optical depth of the molecules. Because N$_2$H$^+$(1-0), HCO$^+$(1-0), HCN(1-0), HNC(1-0) have relatively large optical depth, we just detect these molecules in the outer layer of the molecular cloud. While dust emission at infrared wavelengths is optically thin, we can detect all dust emission along the line of sight, and derive the total H$_2$ column density. Such a trend should be more obvious for molecules with higher optical depth. The optical depth of N$_2$H$^+$(1-0) is much smaller than those of  HCO$^+$(1-0), HCN(1-0), HNC(1-0). Therefore the N$_2$H$^+$ column density we detected is closer to its true value than those of other molecules we studied.  The abundances of N$_2$H$^+$ show relatively small variation.The highest value is $\sim$5 times the lowest value. The corresponding values are $\sim$10 for HCO$^+$, HCN and HNC.

In general, HCN and HCO$^+$ abundances increase as temperature increases, while N$_2$H$^+$ and HNC abundances decrease due to the chemical properties described in Section \ref{sec4.1}. Abundances of all four molecules increase when temperature increases. Considering that the pixels with higher temperature tend to have lower column density, we conclude that the result also could be attributed the fact that the abundances of these molecules are mainly dominated by the H$_2$ column density.

\begin{figure}
\figurenum{6}
  \begin{center}
  \includegraphics[width=0.95\textwidth]{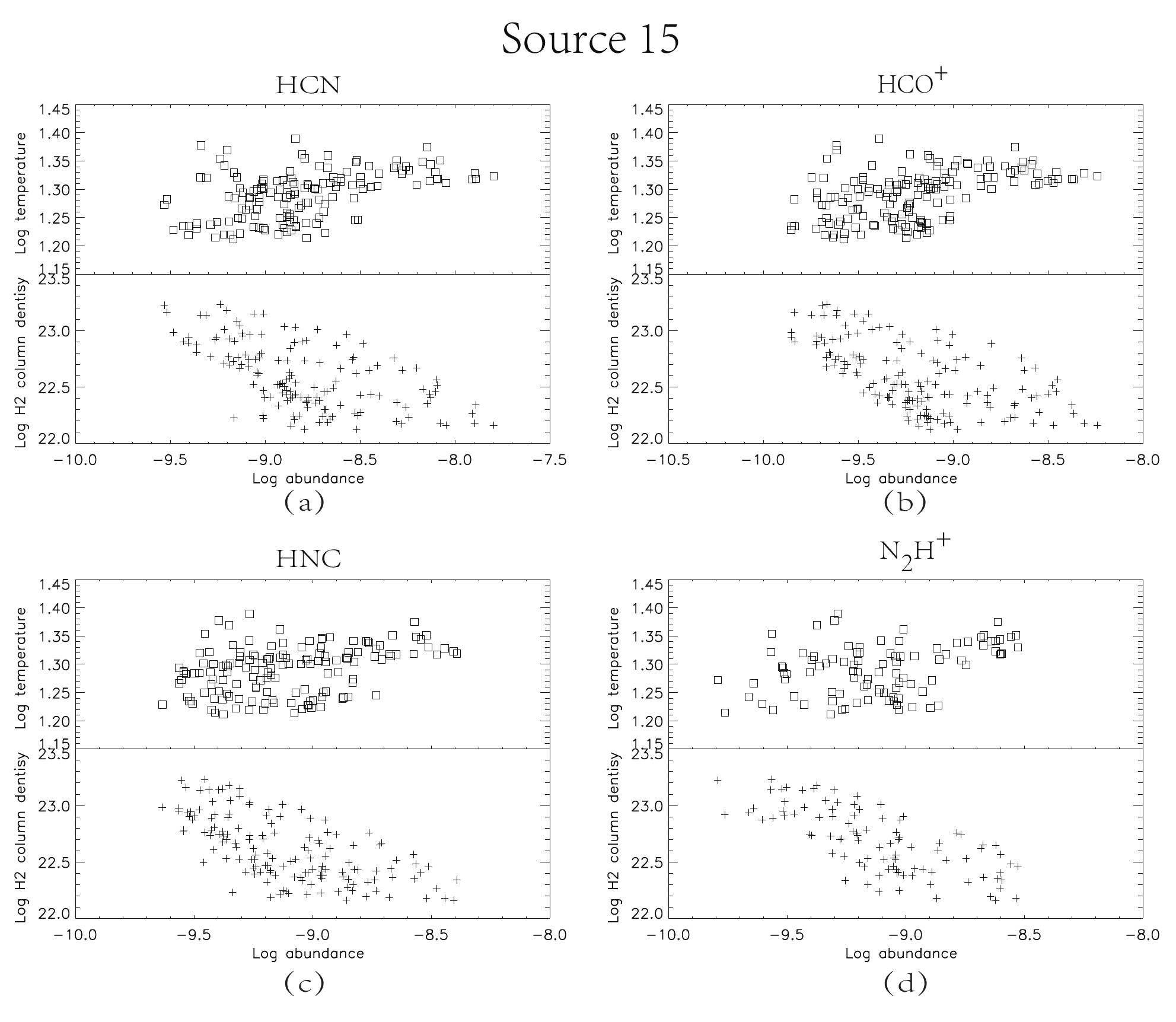}
  \end{center}
  \caption{The abundances of N$_2$H$^+$, HCO$^+$, HCN, HNC as a function of H$_2$ column density and temperature for source 15 (G12.774+00.337), respectively. This source is an example of Group II, it has 20 cm continuum emission, and was classified into the fourth type.}\label{fig6}
\end{figure}

\begin{figure}
\figurenum{7}
  \begin{center}
  \includegraphics[width=0.95\textwidth]{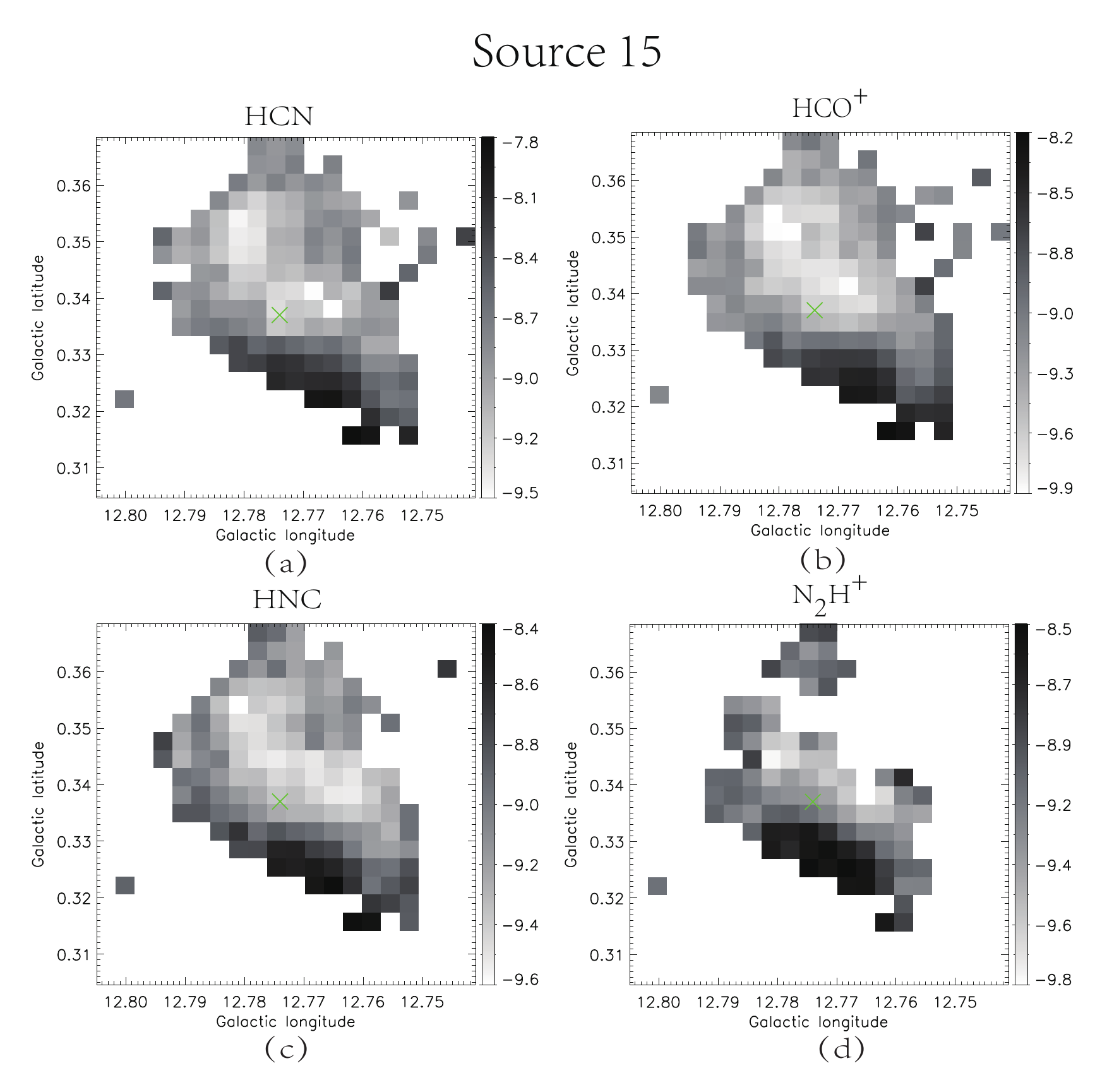}
  \end{center}
  \caption{The abundance distributions of N$_{2}$H$^{+}$, HCO$^{+}$, HCN and HNC of the source 15 (G12.774+00.337) in gray scale. The black bars show the units of in log. This source is an example of Group II, it was classified into the fourth type.}\label{fig7}
\end{figure}

We show source 15 as an example of Group II sources in Figure \ref{fig6}. The abundances of N$_2$H$^+$, HCO$^+$, HCN, HNC increase as H$_2$ column density decreases. This suggests that their abundance variations were also dominated by the H$_2$ column density. However, the dispersion of abundances corresponding to the same H$_2$ column density becomes larger. Such dispersion also increases as H$_2$ column density decreases. This is probably due to the effect of active star formation near it, because the chemistry is sensitive to temperature, density, and ionisation degree. Group II sources are associated with an infrared bubble, which is excited by a group of OB stars. These OB stars affect the temperature and ionisation degree by strong UV radiation, shock and heat. Such effect is more remarkable in the low column density regions where the extinction is small. It can be seen from Figure \ref{fig6}, for the pixels with the same column density, that those have higher abundances tend to have higher temperature. Figure \ref{fig7} shows that the abundance distributions of N$_2$H$^+$, HCO$^+$, HCN, HNC of source 15 are similar to those of source 82. The abundances are low in the center, and high in the edge of the clump. However, the abundance distributions of these four molecules show obvious asymmetry in source 15. All of them reach the maximum values in the southern edge of the clump. This indicates that the clump chemistry was affected by the environment in this direction. The abundance of HCN in the whole map increases very much comparing with that of source 82. This is consistent with the fact source 15 has higher temperature.

\subsection{The spatial variation of the abundance ratios}\label{sec4.3}
\begin{figure}
\figurenum{8}
  \begin{center}
  \includegraphics[width=0.95\textwidth]{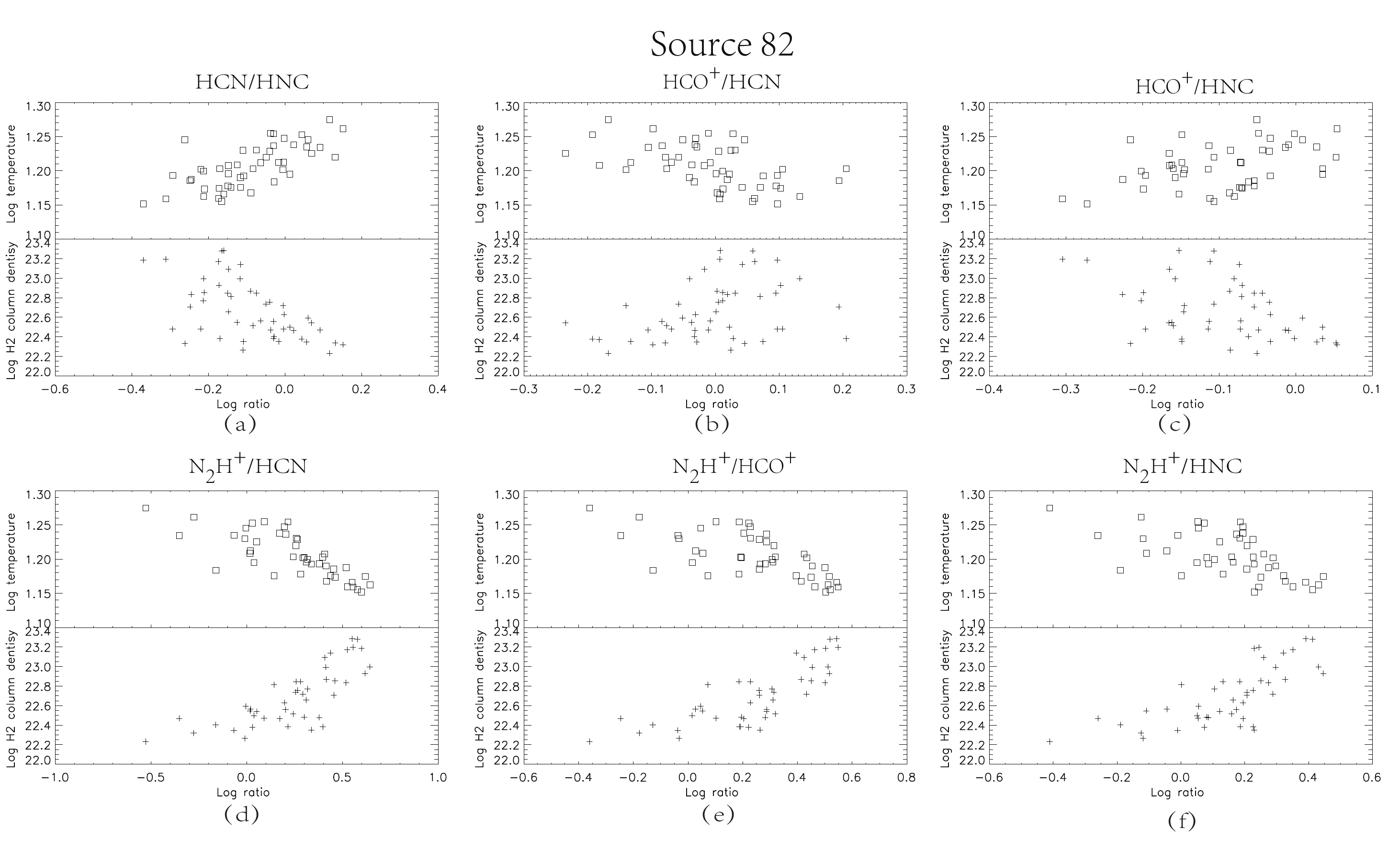}
  \end{center}
  \caption{The abundance ratios HCN/HNC, HCO$^+$/HCN, HCO$^+$/HNC, N$_2$H$^+$/HCN, N$_2$H$^+$/HCO$^+$, and N$_2$H$^+$/HNC as a function of H$_2$ column density and temperature for the source 82 (G354.945-00.539). This source is an example of Group I.}\label{fig8}
\end{figure}

The abundances of N$_2$H$^+$, HCO$^+$, HCN and HNC are affected by H$_2$ column density and environment to a great extent. It is more helpful to study the spatial variations of the abundance ratios between these molecules. We show the source 82 as an example of Group I sources (see Figure \ref{fig8}). The abundance ratio of HCN/HNC shows a similar increasing trend as a function of decreasing H$_2$ column density and increasing temperature. This is consistent with previous results that HNC transfers into HCN at higher temperature,  and the abundance ratio of HCN/HNC is strongly dependent on the temperature \citep{vas11}. The abundance ratios of HNC/HCN, HNC/HCO$^+$, HCO$^+$/HCN, N$_2$H$^+$/HCO$^+$, N$_2$H$^+$/HCN and N$_2$H$^+$/HNC increase as H$_2$ column density increases and temperature decreases. Such results suggest that, increasing H$_2$ column density and decreasing temperature result in the observed decrease in the abundances of (in order of rapidity from highest to lowest) HCN, HCO$^+$, HNC, and N$_2$H$^+$ (see Table \ref{tab5}).

\begin{table*}
\scriptsize
 \centering
  \begin{minipage}{155mm}
   \caption{We show the decrease of four molecules' abundances as a function of N(H$_2$), with (----) being most rapid decrease, and (-) least..}\label{tab5}
    \begin{tabular}{lll}
        \hline
        \multicolumn{1}{c}{Species Name} & \multicolumn{1}{c}{Abundance as a function of N(H$_2$)}\\
        \hline
  HCN       &   - - - - \\
   HCO$^+$ & - - -   &   \\
    HNC  &   - -    \\
  N$_2$H$^+$  & -  \\
      \hline
     \end{tabular}
     \medskip
   \end{minipage}
\end{table*}

\begin{figure}
\figurenum{9}
  \begin{center}
  \includegraphics[width=0.95\textwidth]{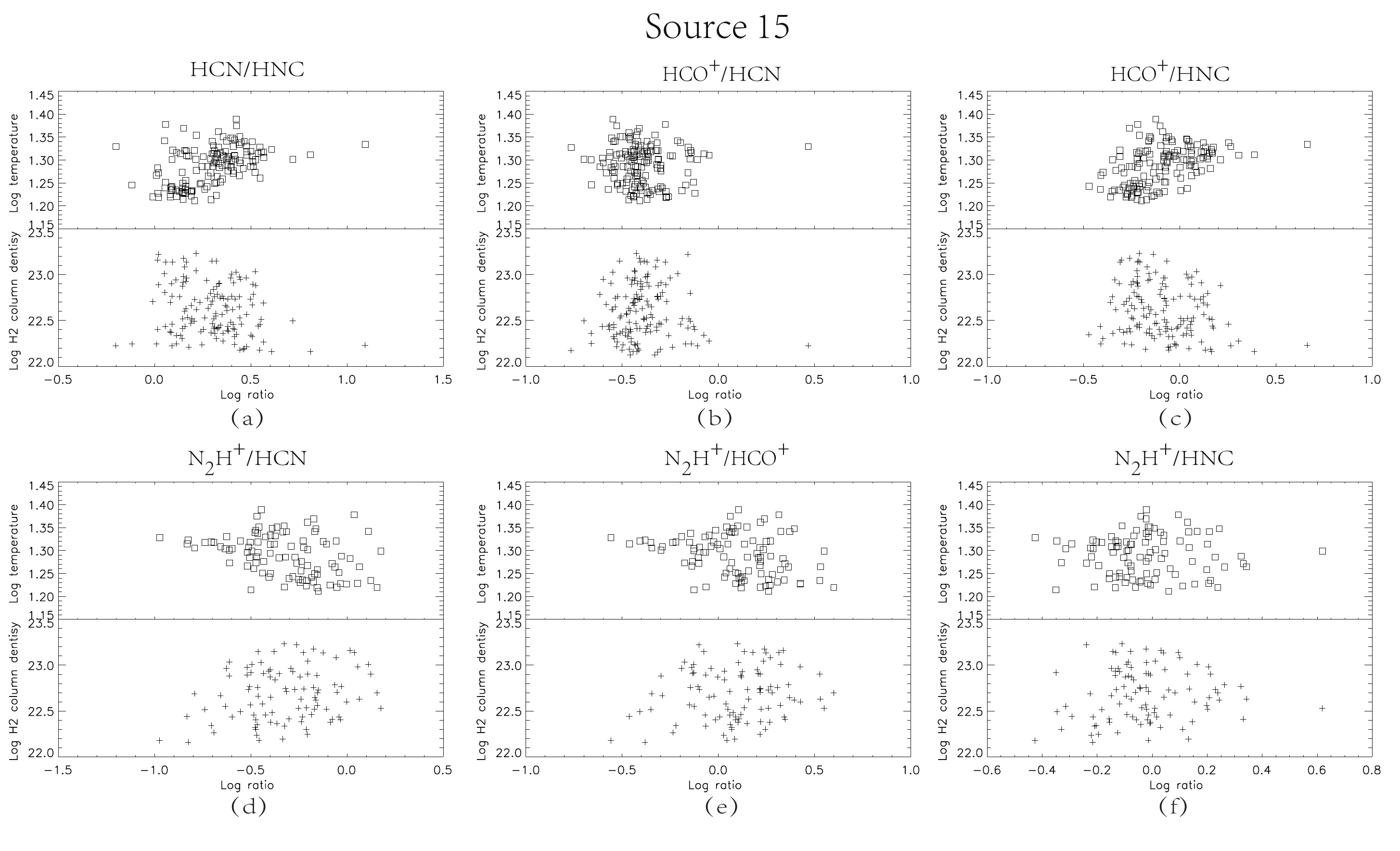}
  \end{center}
  \caption{The abundance ratios HCN/HNC, HCO$^+$/HCN, HCO$^+$/HNC, N$_2$H$^+$/HCN, N$_2$H$^+$/HCO$^+$, and N$_2$H$^+$/HNC as a function of H$_2$ column density and temperature for the source 15 (G12.774+00.337). This source is an example of Group II, it was classified into the fourth type.}\label{fig9}
\end{figure}

We show the source 15 as an example of Group II sources. The abundance ratios of HNC/HCN, HNC/HCO$^+$, HCO$^+$/HCN, N$_2$H$^+$/HCO$^+$, N$_2$H$^+$/HCN and N$_2$H$^+$/HNC show only a weak increasing trend as a function of increasing H$_2$ column density and decreasing temperature (see Figure \ref{fig9}). This is consistent with the fact that their abundances have larger dispersion due to the effect of the environment.

For the sources 82 and 15, those abundance ratios show obvious increasing trends with increasing H$_2$ column density and decreasing temperature. However, this is not true for all Group I and II sources. Though a number of Group I and II sources display similar trends for abundance ratios as the sources 82 and 15 do, many of them do not show clear correlation between the abundance ratios and the H$_2$ column density and temperature (see Figure 8A online). This may be attributed to the complex environment of massive star forming regions.

\subsection{The correlation between the column density of H$_2$ and those of N$_2$H$^+$, HCO$^+$, HCN and HNC}\label{sec4.4}
\begin{figure}
\figurenum{10}
  \begin{center}
  \includegraphics[width=0.95\textwidth]{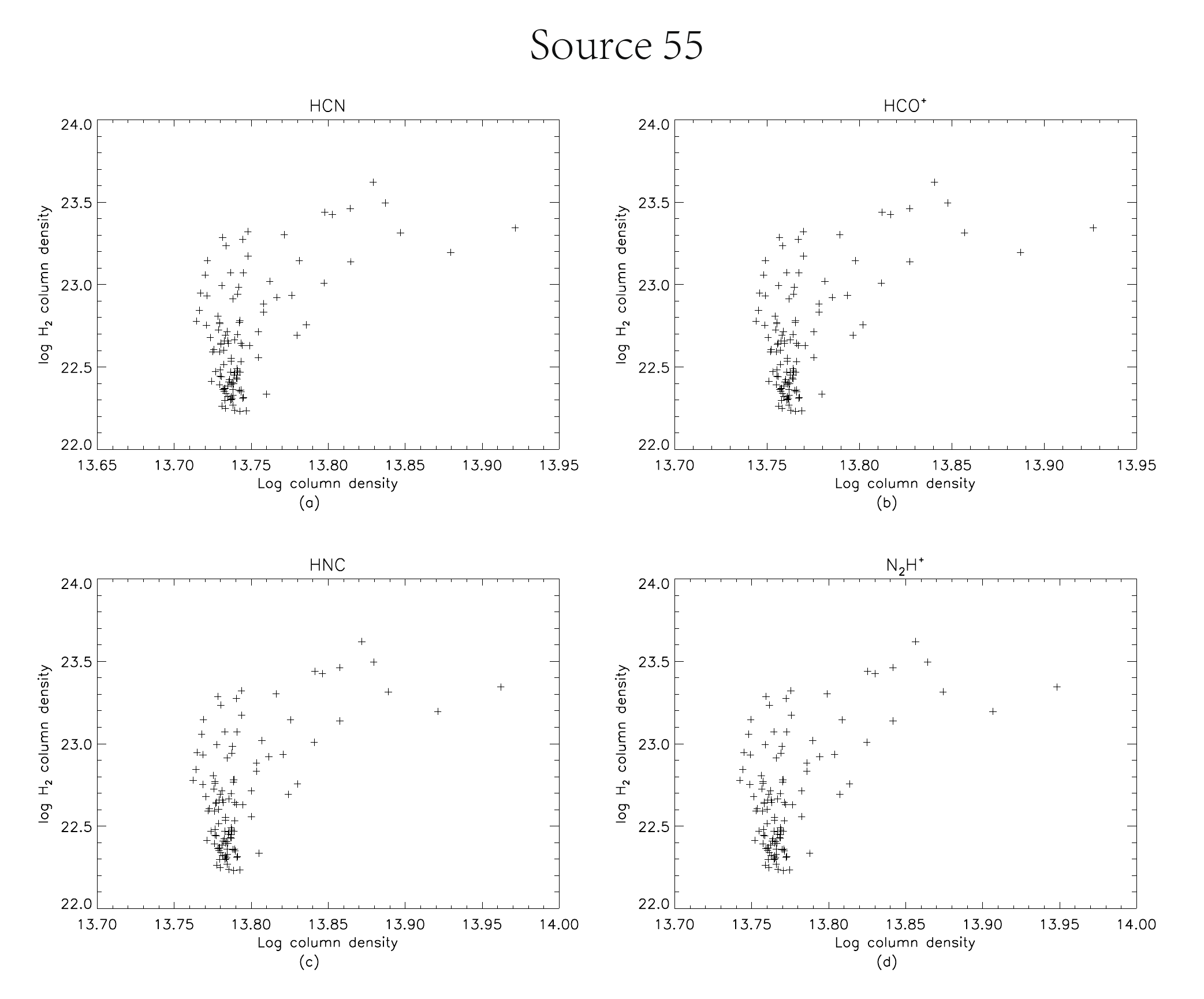}
  \end{center}
  \caption{The H$_2$ column densities as a function of the column densities of N$_2$H$^+$, HCO$^+$, HCN and HNC respectively for the source 55 (G012.48+00.506). This source belongs to Group II. }\label{fig10}
\end{figure}

Figure 6A online displays the correlations between the the H$_2$ and N$_2$H$^+$, HCO$^+$, HCN and HNC column densities pixel-by-pixel for 90 sources. For most sources of our sample (e.g., the source 55 in Figure \ref{fig10}, the correlations between the H$_2$ column density and N$_2$H$^+$, HCO$^+$, HCN and HNC column densities are relatively complex. When the H$_2$ column density is lower than a threshold value, the H$_2$ column density displays an inversely linear correlation with N$_2$H$^+$, HCO$^+$, HCN and HNC column densities respectively. On the other hand, when the H$_2$ column density is higher than the threshold value, the N$_2$H$^+$, HCO$^+$, HCN and HNC column densities tend to be higher. These sources usually overlap with the infrared bubble or contain a small infrared bubble (see Figure 3A online). This indicates that infrared bubble inside the clump or nearby it may affect the abundance of these molecules.

For other sources of our sample (e.g., the source 51 in Figure 6A), the H$_2$ column density displays an obvious decreasing trend as molecules' column densities increase. The correlation is nearly linear. This clearly shows that column densities of N$_2$H$^+$, HCO$^+$, HCN and HNC detected at pixels with low H$_2$ column densities are larger than those detected at pixels with high H$_2$ column densities. Three-color images of these sources (see Figure 3A online) suggest that there is no infrared bubble inside or nearby them.

 Figure 6A clearly shows that, whether there is feedback from star formation or not, the column density variations of N$_2$H$^+$, HCO$^+$, HCN and HNC are much smaller than that of H$_2$ column densities. Their abundances are dominated by H$_2$ column densities.

\subsection{The effect of 20\,cm continuum emission on abundances and abundance ratios}\label{sec4.5}

\begin{table*}
\scriptsize
 \centering
  \begin{minipage}{155mm}
   \caption{We classified 51 sources associated with 20\,cm continuum emission into four types (see Table \ref{tab1}). Here we display the trend of abundances with 20\,cm continuum emission for the four types of sources.}\label{tab6}
    \begin{tabular}{lllll}
        \hline
        \multicolumn{1}{c}{Type} & \multicolumn{1}{c}{Abundance with}\\
         \multicolumn{1}{c}{} & \multicolumn{1}{c}{20cm continuum emission} \\
         \hline
  First       &   decreasing  \\
  Second  &   increasing    \\
  Third    &  decreasing ($\geq$ $\sim$10 mJy/beam)  \\
              &  increasing ($\leq$ $\sim$10 mJy/beam)\\
  Fourth  & no obvious trend  \\
      \hline
     \end{tabular}
     \medskip
   \end{minipage}
\end{table*}

Since 20\,cm continuum emission usually indicates the appearance of OB stars, their feedback may affect the chemistry of nearby molecular clumps. We used 20\,cm continuum emission data of 70 sources from NVSS survey, and convolved and regridded them into the resolution and pixel size of 500$\mu$m data of the Herschel survey. Finally 51 sources have more than 10 pixels overlapped with the 20\,cm continuum emission (see Table \ref{tab1}).  We display the distribution of the 20\,cm continuum emission of these 51 sources in Figure 9A online, and plot the 20\,cm continuum emission flux as a function of abundances of N$_2$H$^+$, HCO$^+$, HCN, and HNC for 51 sources in Figure 10A online, and the 20\,cm continuum emission flux as a function of abundance ratios respectively in Figure 11A online. From an examination of Figure 10A, we found that they could be classified into four types (see Table \ref{tab1}). We simply called them as the first, second, third and fourth type, and discuss their properties respectively in the following.

\subsubsection{The first type}\label{sec4.5.1}
\begin{figure}
\figurenum{11}
  \begin{center}
  \includegraphics[width=0.95\textwidth]{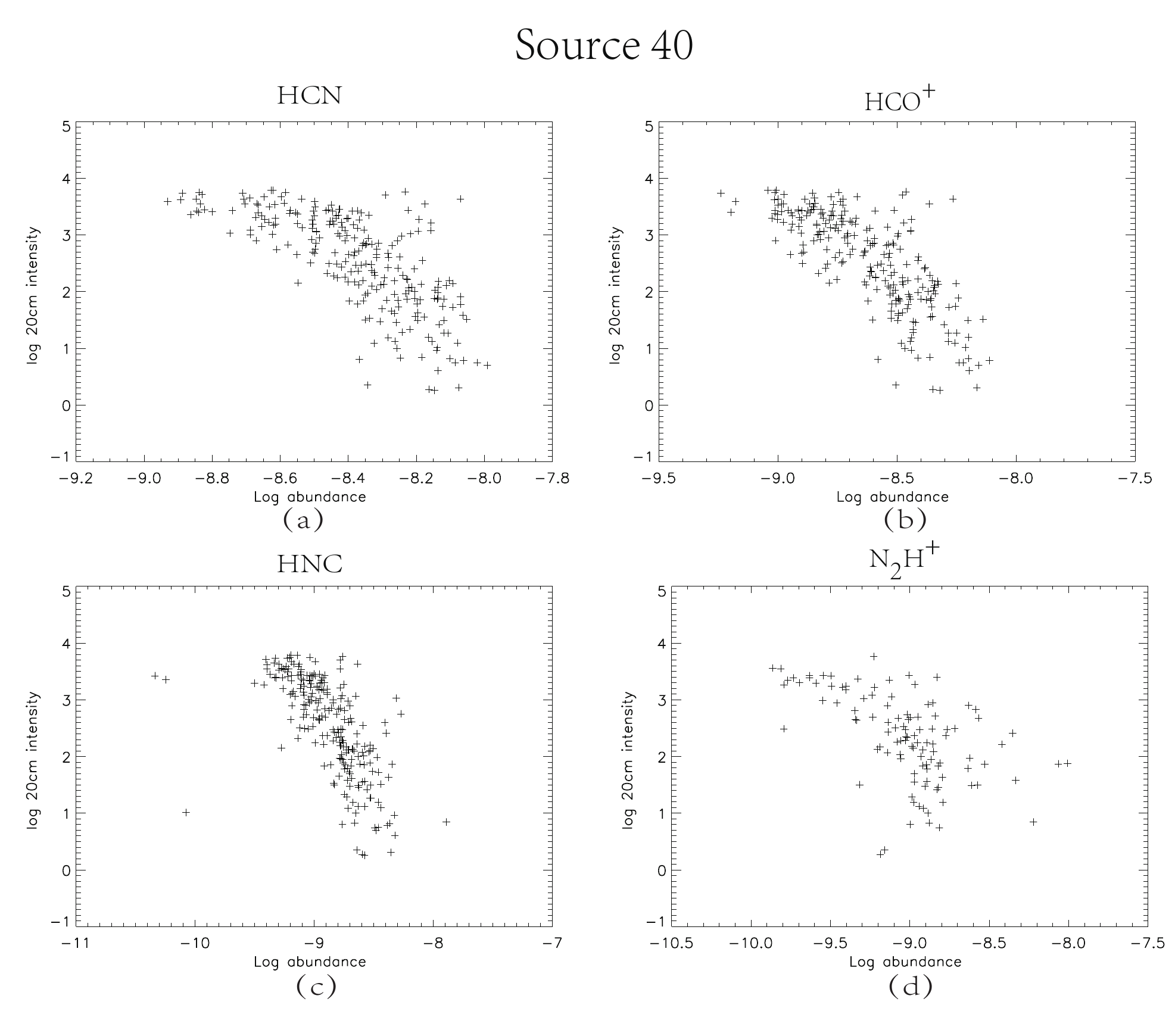}
  \end{center}
  \caption{The abundances of N$_2$H$^+$, HCO$^+$, HCN, and HNC as a function of 20cm continuum emission flux for the source 40 (G353.198+0.927) in log-log. This source is an example of the first type, it belongs to Group II. }\label{fig11}
\end{figure}

\begin{figure}
\figurenum{12}
  \begin{center}
  \includegraphics[width=0.95\textwidth]{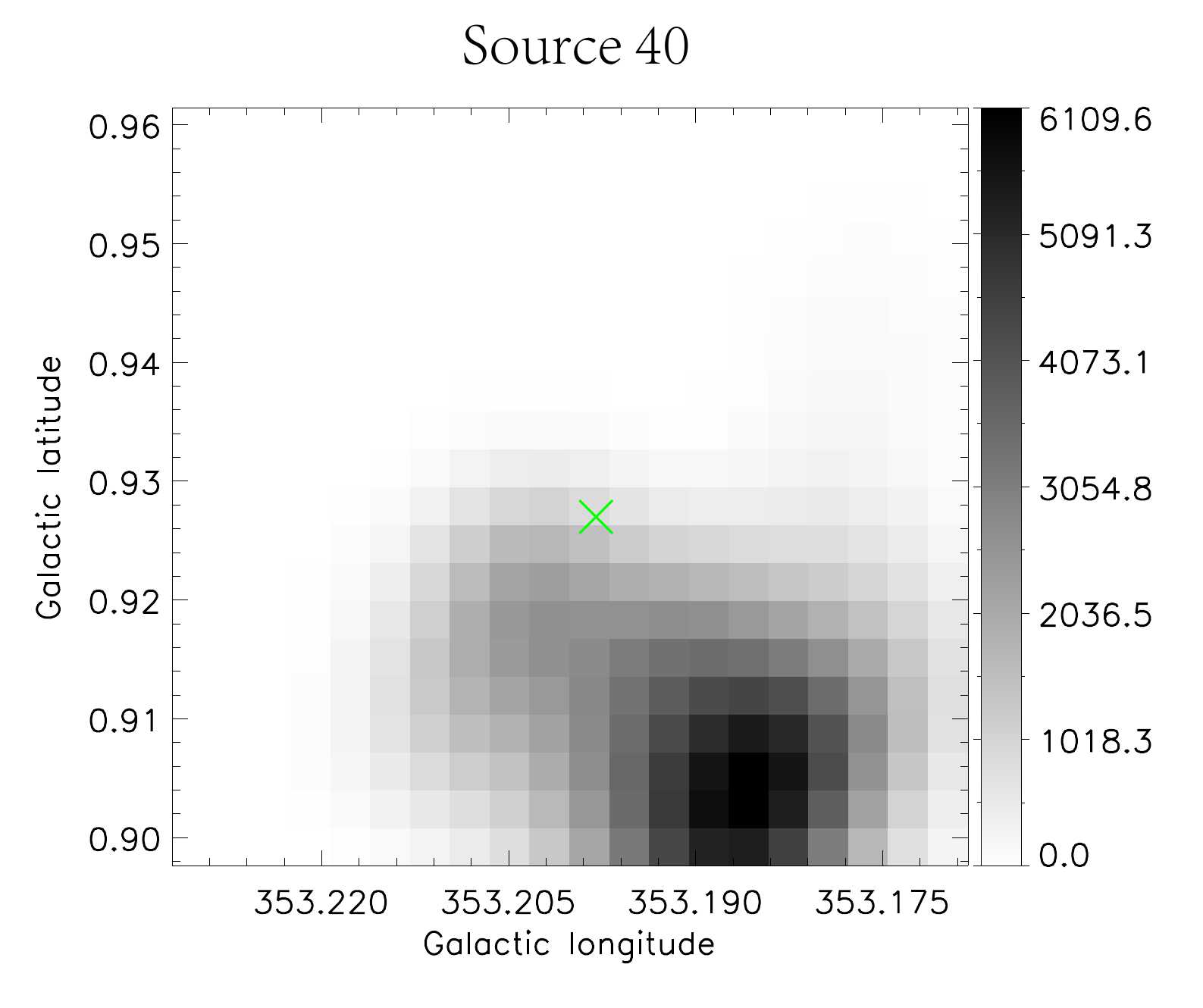}
  \end{center}
  \caption{The distributions of 20cm continuum emission for the Source 40 (G353.198+0.927), the unit of the gray scale bar is mJy/beam. The rms noise $\sigma$$\sim$0.5mJy/beam. The lower limit for the source is 3$\sigma$$\sim$1.5 mJy/beam. This source is an example of the first type, it belongs to Group II.}\label{fig12}
\end{figure}

For the first type, the abundances of N$_2$H$^+$, HCO$^+$, HCN, and HNC decrease as a function of increasing 20\,cm continuum emission flux (see Table \ref{tab6}). We show source 40 as an example in Figure \ref{fig11}. When we check the abundance distributions of N$_2$H$^+$, HCO$^+$, HCN, and HNC of the source 40 in Figure 4A and the distribution of 20cm continuum emission in Figure \ref{fig12}, it is evident that the abundances of HCO$^+$, HCN, and HNC are larger when they are far away from the 20\,cm continuum emission peak. The distribution of N$_2$H$^+$ abundance is much different from that of the other three molecules. It mainly appears in the northern part of the mapping region. The reason is still unclear.

\begin{figure}
\figurenum{13}
  \begin{center}
  \includegraphics[width=0.95\textwidth]{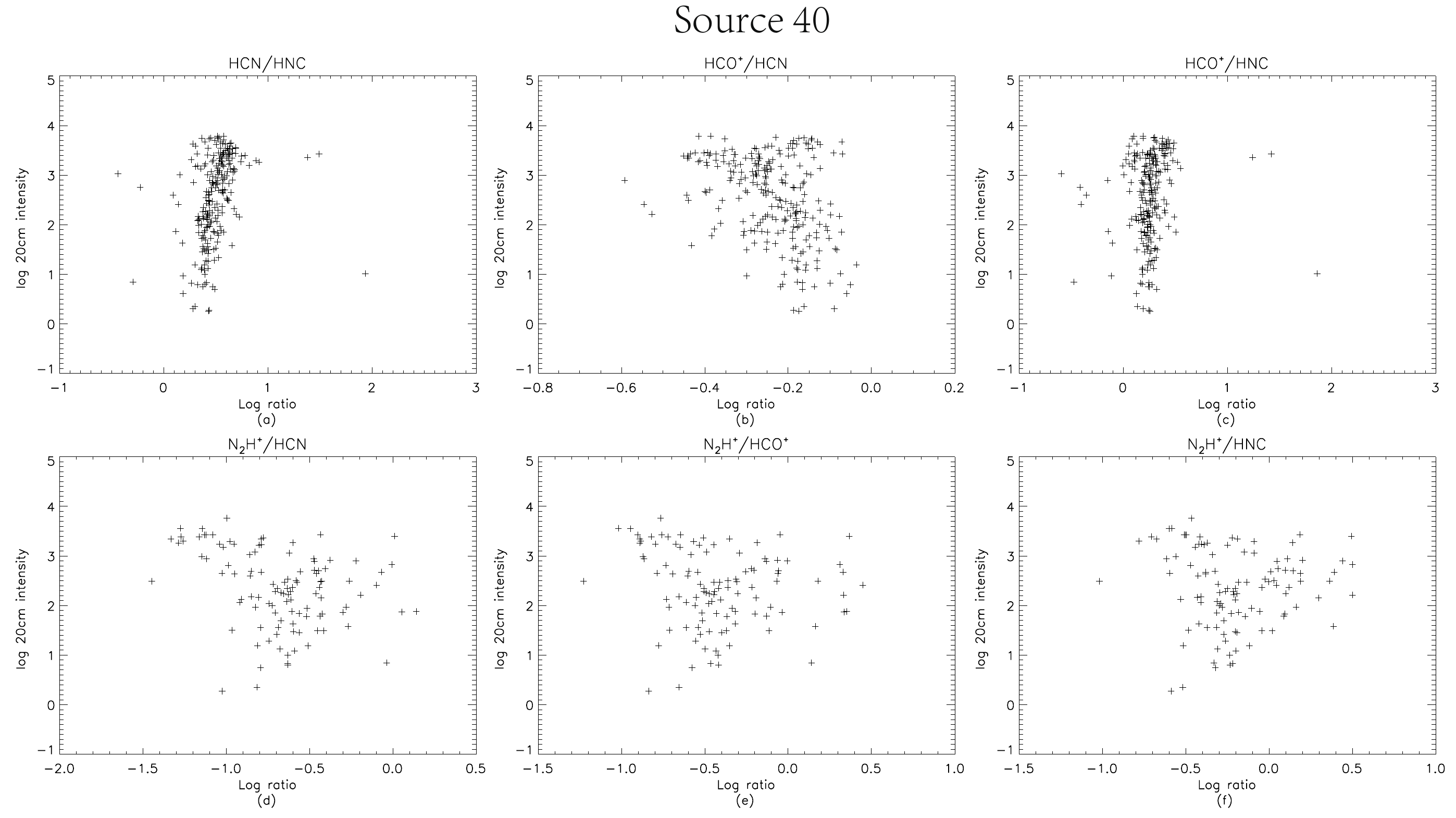}
  \end{center}
  \caption{The abundance ratios among N$_2$H$^+$, HCO$^+$, HCN, and HNC as a function of 20cm continuum emission flux for the source 40 (G353.198+0.927) in log-log. This source is an example of the first type, it belongs to Group II.}\label{fig13}
\end{figure}

Figure \ref{fig13} displays the abundance ratios between  N$_2$H$^+$, HCO$^+$, HCN, and HNC as a function of 20\,cm continuum emission flux. Only abundance ratio of N$_2$H$^+$/HNC shows a weak trend of rising as 20\,cm continuum emission flux decreases. This indicates that N$_2$H$^+$ abundance decreases more rapidly when it fronts onto strong 20\,cm continuum emission. The other abundance ratios, however, do not show obvious trend as 20\,cm continuum emission flux decreases.

\begin{figure}
\figurenum{14}
  \begin{center}
  \includegraphics[width=0.95\textwidth]{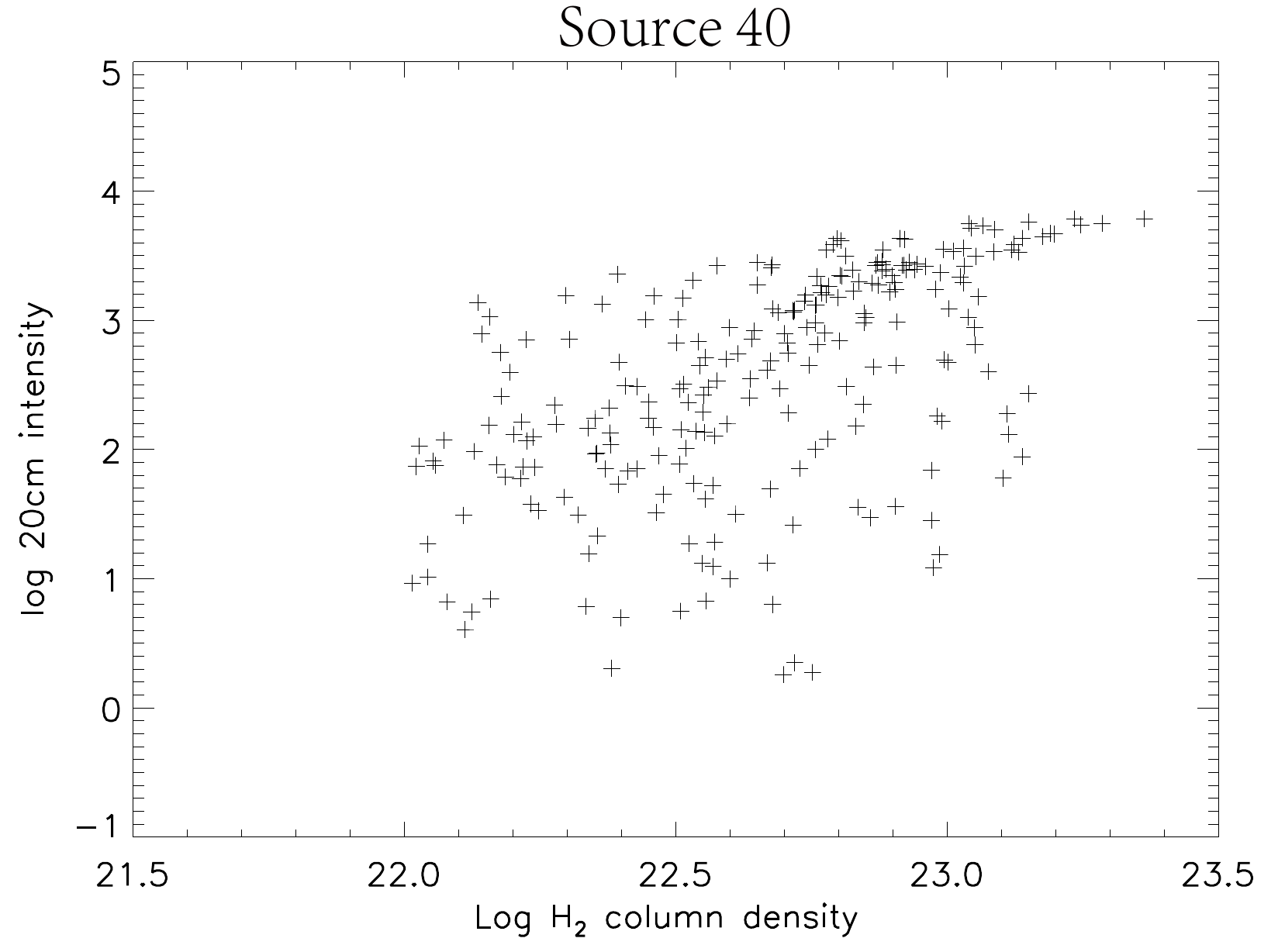}
  \end{center}
  \caption{The H$_2$ column density as a function of 20cm continuum emission flux for the source 40 (G353.198+0.927) in log-log. This source is an example of the first type, it belongs to Group II.}\label{fig14}
\end{figure}

For the source 40, 20\,cm continuum emission flux has a positively proportional correlation with H$_2$ column density, though the correlation is not obvious (see Figure \ref{fig14}, in Figure 12A online we show the 20\,cm continuum emission flux as a function of H$_2$ column density for 51 sources). Considering that abundances of N$_2$H$^+$, HCO$^+$, HCN, and HNC have an inversely proportional correlation with the H$_2$ column density, the abundance variations of these four molecules with the 20\,cm continuum emission could also be attributed to the variation of H$_2$ column density to a great extent.

\subsubsection{The second type}\label{sec4.5.2}
\begin{figure}
\figurenum{15}
  \begin{center}
  \includegraphics[width=0.95\textwidth]{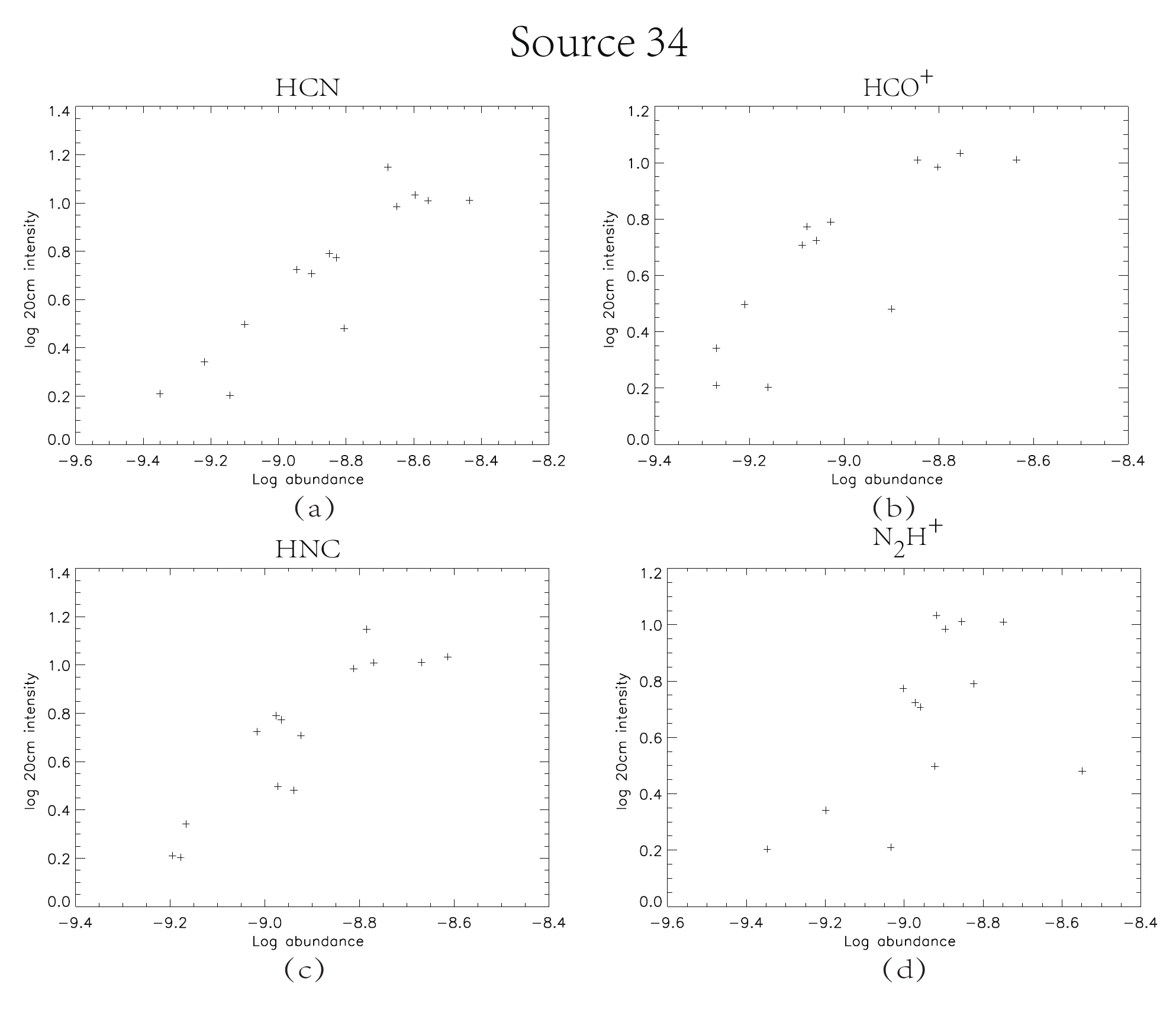}
  \end{center}
  \caption{The abundances of N$_2$H$^+$, HCO$^+$, HCN, and HNC as a function of 20 cm continuum emission flux for the source 34 (G352.684-0.120) in log-log. This source is an example of the second type, it belongs to Group I.}\label{fig15}
\end{figure}

For the second type of sources, the abundances of N$_2$H$^+$, HCO$^+$, HCN, and HNC increase as a function of increasing 20\,cm continuum emission flux (see Table \ref{tab6}). We show source 34 as an example in Figure \ref{fig15}. By checking the abundance distributions of these four molecules in Figure 4A and 20\,cm continuum emission in Figure 9A, we found that the abundances of these molecules are usually small at the center of the clump, and are large at the edges of the clump, but are much larger in the eastern edge of the clump, which fronts onto the 20\,cm continuum emission.

\begin{figure}
\figurenum{16}
  \begin{center}
  \includegraphics[width=0.95\textwidth]{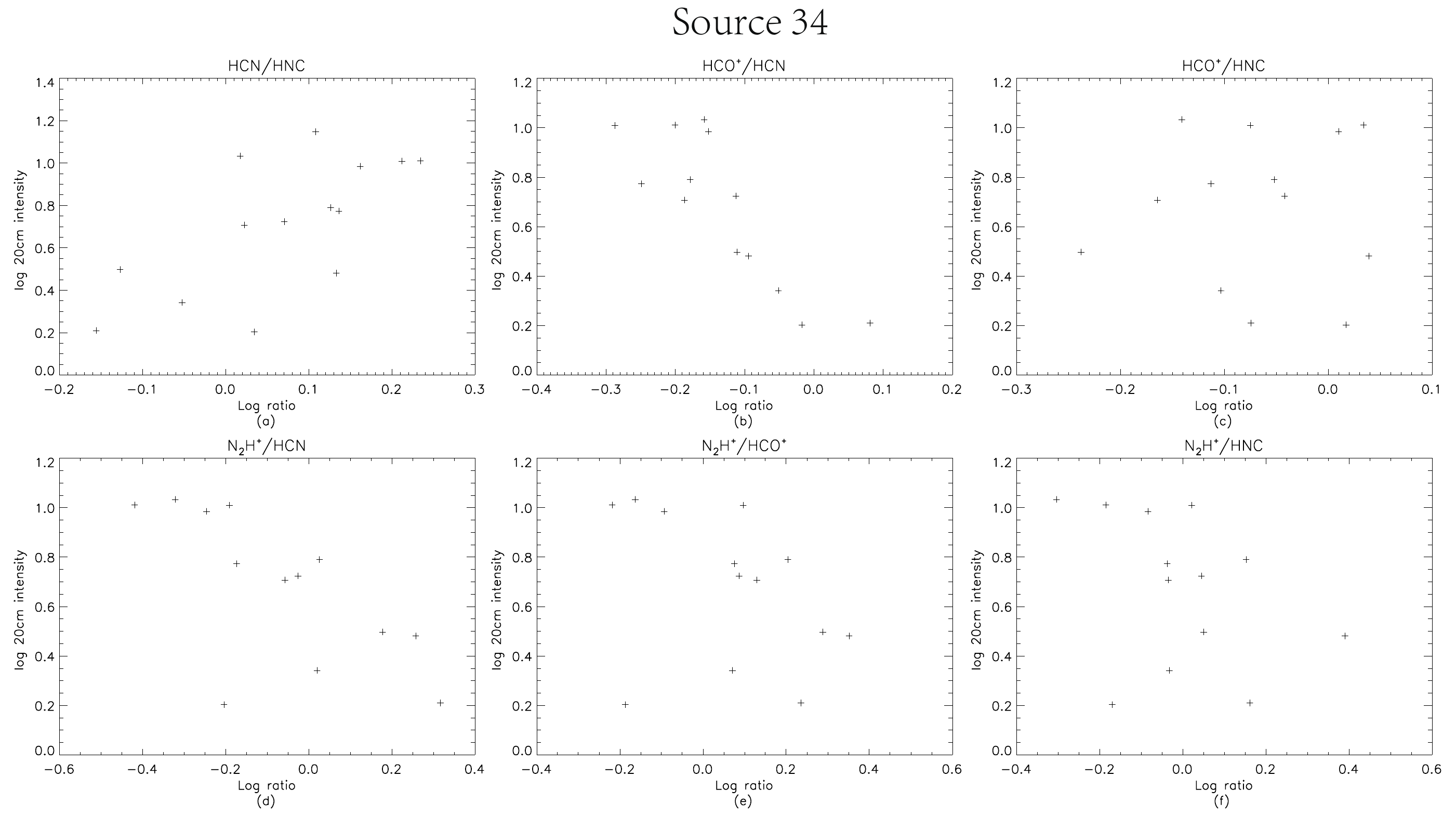}
  \end{center}
  \caption{The abundance ratios among N$_2$H$^+$, HCO$^+$, HCN, and HNC as a function of 20cm continuum emission flux for the source 34 (G352.684-0.120) in log-log. This source is an example of the second type, it belongs to Group I.}\label{fig16}
\end{figure}

Figure \ref{fig16} displays the the abundance ratios as a function of 20\,cm continuum emission flux, where abundance ratios of N$_2$H$^+$/HCN and HCO$^+$/HCN show a weak increasing trend as 20\,cm continuum emission flux decreases. The results suggest that HCN abundance increases more rapidly than N$_2$H$^+$ and HCO$^+$ do. When 20\,cm continuum emission becomes strong. N$_2$H$^+$/HNC also displays a very weak increasing trend as 20\,cm continuum emission flux decreases. This indicates that HNC abundance decreases more rapidly than N$_2$H$^+$ does when 20\,cm continuum emission becomes strong. The other three abundance ratios show no obvious variation as a function of 20\,cm continuum emission flux.

For the source 34, 20\,cm continuum emission flux shows an inversely proportional correlation with H$_2$ column density (see Figure 12A). Considering that abundances of N$_2$H$^+$, HCO$^+$, HCN, and HNC have an inversely correlation with the H$_2$ column density, the abundance variations of these four molecules should be positively proportional with the 20\,cm continuum emission. Their variations with the 20\,cm continuum emission in this case could also be attributed to the variation of H$_2$ column density to a great extent. However, the correlation between 20\,cm continuum emission and abundance is not a simple reverse of the correlation between 20\,cm continuum emission and H$_2$ column density. This indicates that the environment such as UV radiation traced by 20\,cm continuum emission may contribute to the abundance variations of these four molecules. More works based on chemical  model are needed to confirm this.

\subsubsection{The third type}\label{sec4.5.3}
We noted that the 20\,cm continuum emission flux of the second type of sources is much smaller than that of the first type sources. This suggests the abundances of N$_2$H$^+$, HCO$^+$, HCN, and HNC have an inversely proportional correlation with the 20\,cm continuum emission flux when it is very strong, and have a positively proportional correlation with the 20\,cm continuum emission flux when it is weak.

\begin{figure}
\figurenum{17}
  \begin{center}
  \includegraphics[width=0.95\textwidth]{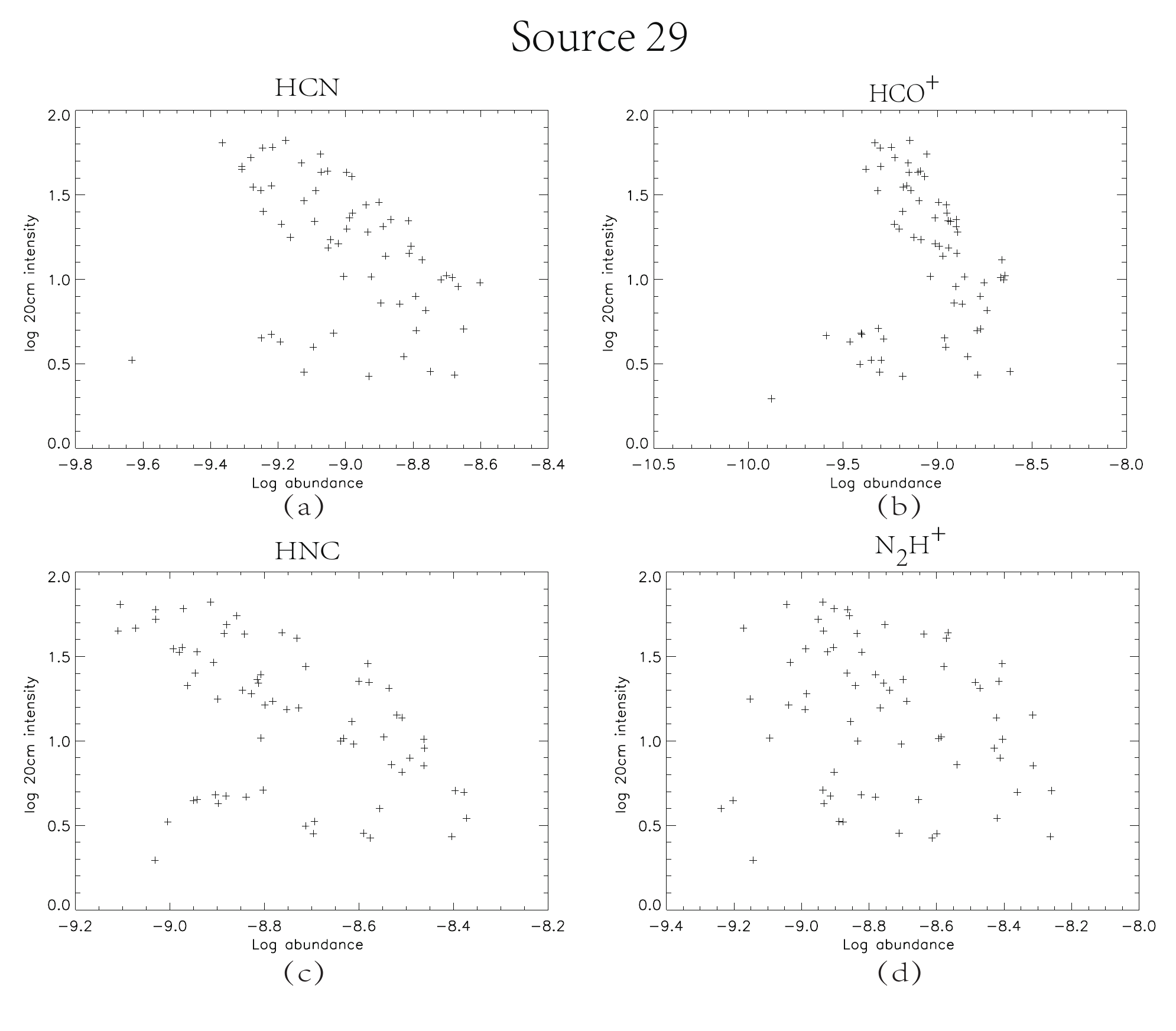}
  \end{center}
  \caption{The abundances of N$_2$H$^+$, HCO$^+$, HCN, and HNC as a function of 20cm continuum emission flux for the source 29 (G351.040-0.336) in log-log. This source is an example of the third type, it belongs to Group II.}\label{fig17}
\end{figure}

As for the third type of sources, the abundances of N$_2$H$^+$, HCO$^+$, HCN, and HNC increase as a function of 20\,cm continuum emission flux when it is less than a certain value, and decrease as a function of 20\,cm continuum emission flux when it is greater than the same value (see Table \ref{tab6}). We show source 29 as an example in Figure \ref{fig17}. When 20\,cm continuum emission flux is greater than $\simeq$10 mJy/beam, HCO$^+$, HCN, and HNC abundances display an obvious increasing trend as 20\,cm continuum emission flux decreases. And when the 20\,cm continuum emission flux is less than $\simeq$10 mJy/beam, these abundances display a weak increasing trend as 20\,cm continuum emission flux increases. The three color image of the source indicates that new stars are being formed in its center (Figure 3A), and the 20\,cm continuum emission is stronger in northern part of the clump (Figure 9A). N$_2$H$^+$, HCO$^+$, HCN, and HNC abundances are small in the central part of the clump and eastern and southern edges (see Figure 4A). These results show that abundances are large in pixels where 20\,cm continuum emission is relatively strong and column density is relatively low. N$_2$H$^+$ abundance displays a similar trend of variation when 20\,cm continuum emission flux is greater or less than $\simeq$10 mJy/beam, but the dispersion is larger.

\begin{figure}
\figurenum{18}
  \begin{center}
  \includegraphics[width=0.95\textwidth]{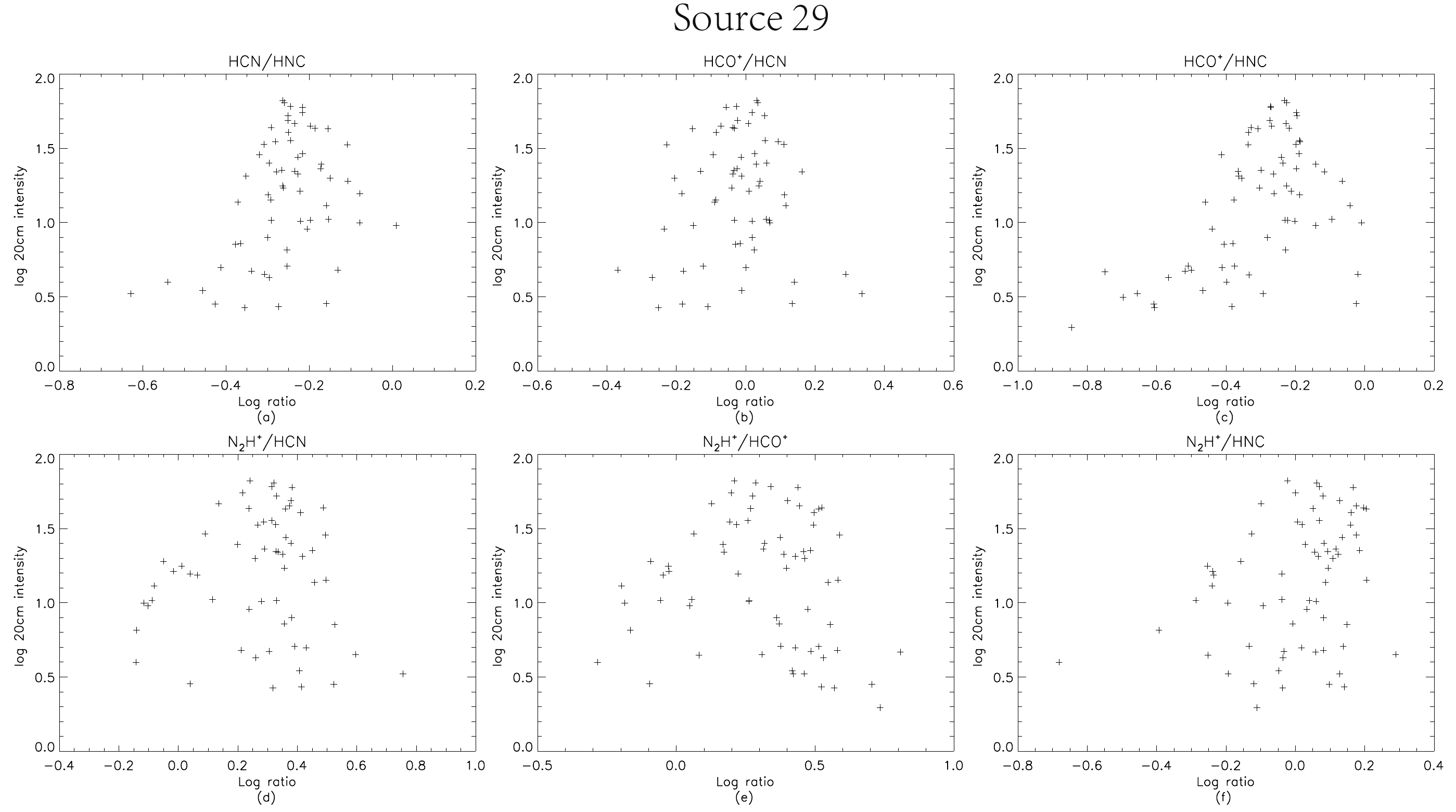}
  \end{center}
  \caption{The abundance ratios among N$_2$H$^+$, HCO$^+$, HCN, and HNC as a function of 20cm continuum emission flux for the source 29 (G351.040-0.336) in log-log. This source is an example of the third type, it belongs to Group II.}\label{fig18}
\end{figure}

The abundance ratios of N$_2$H$^+$/HCN, N$_2$H$^+$/HCO$^+$, N$_2$H$^+$/HNC display a weak increasing trend as a function of increasing 20\,cm continuum emission flux (see Figure \ref{fig18}), and so abundances of HCN, HCO$^+$ and HNC increase more slowly as a function of 20\,cm continuum emission than   N$_2$H$^+$ does. The fact that abundance ratios HCN/HNC, HCO$^+$/HCN, HCO$^+$/HNC show no obvious trend as a function of 20\,cm continuum emission indicates the abundances of HCN, HCO$^+$ and HNC vary at a similar rate in this case.

For the source 29, 20\,cm continuum emission flux shows a positively proportional correlation with H$_2$ column density when it is greater than 10 mJy/beam, and shows an inversely proportional correlation with the H$_2$ column density when it is less than 10 mJy/beam (see Figure 12A). This is contrary to the abundance variations (Figure \ref{fig17}). Based on the same consideration as that for the sources 40 and 34, the abundance variations of N$_2$H$^+$, HCO$^+$, HCN, and HNC with the 20\,cm continuum emission could also be attributed to the variation of H$_2$ column density to a great extent.

 Considering that 20\,cm continuum emission is positively correlated with the UV radiation of nearby OB stars, it is natural to predict the 20\,cm continuum emission is correlated with the spatial variations of the abundance of N$_2$H$^+$, HCO$^+$, HCN, and HNC. However, the abundance variations are likely to be a more complex function of environment and radiative transfer (ie, a combination of opacity, excitation, and radiation field) \citep{sch17}. This may explain the reason for which we did not see obvious correlation between 20cm continuum emission and molecule abundances for many sources, which we described as the fourth type sources (see Table \ref{tab1}, Table \ref{tab6}).

 In a brief, we found good correlations between the molecules' abundances and 20\,cm continuum emission in first, second and third types of sources, which could also be attributed to the variation of H$_2$ column density. We did not find obvious correlations between the abundance ratios and 20\,cm continuum emission.

\section{Conclusions}\label{sec5}
We derived the distribution of abundances of N$_2$H$^+$, HCO$^+$, HCN and HNC and abundance ratios between them, and studied their spatial variations and correlations with H$_2$ column density, temperature and 20cm continuum emission. Our main conclusions are the following:
\begin{enumerate}
 \item We derived the distribution of temperature, H$_2$ column density, abundances of N$_2$H$^+$, HCO$^+$, HCN and HNC and abundance ratio between them for 90 massive star forming clumps.

\item The abundances of N$_2$H$^+$, HCO$^+$, HCN and HNC increase as a function of decreasing column density and increasing temperature. The reason seems to be that these molecules are not optically thin. We just detected them in the outer layer of the molecular cloud, while dust emission is optically thin, we thus derived the total H$_2$ column density along the line of sight. The column densities of these four molecules vary little when the H$_2$ column density vary about one order of magnitude. So the spatial variations of the abundances of these four molecules were dominated by H$_2$ column density.

\item The abundance ratios between N$_2$H$^+$, HCO$^+$, HCN, and HNC also display systemic variations as a function of column density. This could be explained by the chemical properties of these four molecules. However, for the sources overlapped with the bubbles, the abundance ratios between these molecules usually show no obvious trend as a function of column density and temperature. This may be attributed to the effect from the environment.

\item For the sources where the 20cm continuum emission is very strong, the abundances of these four molecules decrease as a function of increasing 20cm continuum emission flux. However, for the sources where the 20cm continuum emission flux is weak, the abundances of these four molecules increase as a function of increasing 20\,cm continuum emission flux. These variations are mainly dominated by H$_2$ column density.

\item The 20\,cm continuum emission is correlated with the UV radiation of OB stars nearby, and so it is probably associated with the spatial abundance variations of the molecules. However, the abundance ratios between these four molecules usually do not show obvious trend as a function of 20\,cm continuum emission flux. This suggests that the effect from the environment that traced by 20\,cm continuum emission do not affect their abundance too much.

\item The spatial variations of the abundances of N$_2$H$^+$, HCO$^+$, HCN and HNC are dominated by H$_2$ column density, and massive stars form in clumps by cluster. As the result, there exists obvious column density variation in massive star forming regions, and therefore obvious spatial variations of chemical properties. It is difficult to extract the chemical properties characterizing the evolution of massive star forming regions.
\end{enumerate}

It should be noted that we use simple approaches to estimate the abundances, such as assuming a fixed excitation temperature, and assuming the local dust temperature to be the gas excitation temperature. In fact, abundance variations are likely to be a more complex function of environment and radiative transfer (i. e., a combination of opacity, excitation, and radiation field) \citet{sch17}. So more works based on chemical models and optical molecular lines are needed.

\section{Appendix}

Captions of 12 Figures online.

Figure 1A: The distributions of temperature for 90 massive star forming clumps, where the black bars show temperature in K.

Figure 2A: The distributions of H$_2$ column density for 90 massive star forming clumps.  The black bars show the logarithm of column density in cm$^{-2}$.

Figure 3A: Three-color images of 90 massive star forming clumps. We show 3.6$\mu$m in blue, 8$\mu$m in green and 24$\mu$m in red.

Figure 4A: The abundance distributions of N$_2$H$^+$, HCO$^+$, HCN and HNC for 90 massive star forming clumps. The black bars show abundances in logarithm.

Figure 5A: The abundance ratio distributions of HCN/HNC, HCO$^+$/HCN, HCO$^+$/HNC, N$_2$H$^+$/HCN, N$_2$H$^+$/HCO$^+$ and N$_2$H$^+$/HNC for 90 massive star forming clumps. The black bars show ratios in logarithm.

Figure 6A: The H$_2$ column densities as a function of the column densities of N$_2$H$^+$, HCO$^+$, HCN and HNC respectively for 90 massive star forming clumps.

Figure 7A: The abundances of N$_2$H$^+$, HCO$^+$, HCN and HNC as a function of H$_2$ column density and temperature respectively for 90 massive star forming clumps.

Figure 8A: The abundance ratios of HCN/HNC, HCO$^+$/HCN, HCO$^+$/HNC, N$_2$H$^+$/HCN, N$_2$H$^+$/HCO$^+$ and N$_2$H$^+$/HNC as a function of H$_2$ column density and temperature respectively for 90 massive star forming clumps.

Figure 9A: The distributions of 20cm continuum emission for 51 massive star forming clumps, the unit of the gray scale bar is mJy/beam. The rms noise $\sigma$$\sim$0.5mJy/beam. The lower limit for each source is 3$\sigma$$\sim$1.5 mJy/beam.

Figure 10A: The abundances of N$_2$H$^+$, HCO$^+$, HCN, and HNC as a function of 20cm continuum emission flux respectively for 51 massive star forming clumps.

Figure 11A: The abundance ratios of HCN/HNC, HCO$^+$/HCN, HCO$^+$/HNC, N$_2$H$^+$/HCN, N$_2$H$^+$/HCO$^+$ and N$_2$H$^+$/HNC as a function of 20cm continuum emission flux respectively for 51 massive star forming clumps.

Figure 12A: The H$_2$ column density as a function of 20cm continuum emission flux for 51 massive star forming clumps.

\section*{Acknowledgments}
This work was funded by The National Natural Science foundation of China under Grant Nos 11433008, 11603063, 11703074, 11703073 and 11763007 and the CAS ¡°Light of West China¡± Program under Grant Nos 2018-XBQNXZ-B-024, 2016-QNXZ-B-23 and 2016-QNXZ-B-22.

\end{document}